\documentclass[letterpaper,twocolumn,prl,aps,superscriptaddress,amsmath,amssymb,floatfix]{revtex4-1}
\usepackage{mathptmx}
\usepackage[latin9]{inputenc}

\usepackage{color}
\usepackage{verbatim}
\usepackage{float}
\usepackage{amsmath}
\usepackage{amssymb}
\usepackage{graphicx}
\usepackage{esint}
\usepackage[T1]{fontenc}
\usepackage[unicode=true,
 bookmarks=true,bookmarksnumbered=false,bookmarksopen=false,
 breaklinks=false,pdfborder={0 0 1},backref=false,colorlinks=true]
 {hyperref}
\hypersetup{
 linkcolor=magenta,urlcolor=blue,citecolor=blue,pdfstartview={FitH},hyperfootnotes=false}

\makeatletter

\usepackage{textcomp}
\usepackage{epstopdf}

\pdfpageheight\paperheight
\pdfpagewidth\paperwidth

\@ifundefined{textcolor}{}{%
 \definecolor{BLACK}{gray}{0}
 \definecolor{WHITE}{gray}{1}
 \definecolor{RED}{rgb}{1,0,0}
 \definecolor{GREEN}{rgb}{0,1,0}
 \definecolor{BLUE}{rgb}{0,0,1}
 \definecolor{CYAN}{cmyk}{1,0,0,0}
 \definecolor{MAGENTA}{cmyk}{0,1,0,0}
 \definecolor{YELLOW}{cmyk}{0,0,1,0}
}

\usepackage{xcolor}
\usepackage{soul}
\definecolor{blue}{rgb}{0,0,1}
\definecolor{red}{rgb}{1,0,0}
\definecolor{green}{rgb}{0,1,0}

\usepackage{soul}

\makeatother

\begin{document}

\affiliation{Institute of Quantum Information and Technology, Nanjing University of Posts and Telecommunications, Nanjing 210003, China}
\affiliation{CAS Key Laboratory of Quantum Information, University of Science and Technology of China, Hefei 230026, China}
\affiliation{Broadband Wireless Communication and Sensor Network Technology, Key Lab of Ministry of Education, Nanjing University of Posts and Telecommunications, Nanjing 210003, China}
\affiliation{CAS Center for Excellence in Quantum Information and Quantum Physics, University of Science and Technology of China, Hefei 230026, China}

\title{Proposal of a free-space-to-chip pipeline for transporting single atoms}
\author{Aiping Liu}
\affiliation{Institute of Quantum Information and Technology, Nanjing University of Posts and Telecommunications, Nanjing 210003, China}
\affiliation{Broadband Wireless Communication and Sensor Network Technology, Key Lab of Ministry of Education, Nanjing University of Posts and Telecommunications, Nanjing 210003, China}
\author{Jiawei Liu}
\affiliation{Institute of Quantum Information and Technology, Nanjing University of Posts and Telecommunications, Nanjing 210003, China}
\affiliation{Broadband Wireless Communication and Sensor Network Technology, Key Lab of Ministry of Education, Nanjing University of Posts and Telecommunications, Nanjing 210003, China}
\author{Zhanfei Kang}
\affiliation{Institute of Quantum Information and Technology, Nanjing University of Posts and Telecommunications, Nanjing 210003, China}
\affiliation{Broadband Wireless Communication and Sensor Network Technology, Key Lab of Ministry of Education, Nanjing University of Posts and Telecommunications, Nanjing 210003, China}

\author{Guang-Jie Chen}
\affiliation{CAS Key Laboratory of Quantum Information, University of Science and
Technology of China, Hefei 230026, China}
\affiliation{CAS Center for Excellence in Quantum Information and Quantum Physics,
University of Science and Technology of China, Hefei 230026, China}
\author{Xin-Biao Xu}
\affiliation{CAS Key Laboratory of Quantum Information, University of Science and
Technology of China, Hefei 230026, China}
\affiliation{CAS Center for Excellence in Quantum Information and Quantum Physics,
University of Science and Technology of China, Hefei 230026, China}
\author{Xifeng Ren}
\affiliation{CAS Key Laboratory of Quantum Information, University of Science and
Technology of China, Hefei 230026, China}
\affiliation{CAS Center for Excellence in Quantum Information and Quantum Physics,
University of Science and Technology of China, Hefei 230026, China}
\author{Guang-Can Guo}
\affiliation{CAS Key Laboratory of Quantum Information, University of Science and
Technology of China, Hefei 230026, China}
\affiliation{CAS Center for Excellence in Quantum Information and Quantum Physics,
University of Science and Technology of China, Hefei 230026, China}

\author{Qin Wang}
\email{qinw@njupt.edu.cn}
\affiliation{Institute of Quantum Information and Technology, Nanjing University of Posts and Telecommunications, Nanjing 210003, China}
\affiliation{Broadband Wireless Communication and Sensor Network Technology, Key Lab of Ministry of Education, Nanjing University of Posts and Telecommunications, Nanjing 210003, China}

\author{Chang-Ling Zou}
\email{clzou321@ustc.edu.cn}
\affiliation{CAS Key Laboratory of Quantum Information, University of Science and
Technology of China, Hefei 230026, China}
\affiliation{CAS Center for Excellence in Quantum Information and Quantum Physics,
University of Science and Technology of China, Hefei 230026, China}

\date{\today}

\begin{abstract}
A free-space-to-chip pipeline is proposed to efficiently transport single atoms from a magneto-optical trap to an on-chip evanescent field trap. Due to the reflection of the dipole laser on the chip surface, the conventional conveyor belt approach can only transport atoms close to the chip surface but with a distance of about one wavelength, which prevents efficient interaction between the atom and the on-chip waveguide devices. Here, based on a two-layer photonic chip architecture, a diffraction beam of the integrated grating with an incident angle of the Brewster angle is utilized to realize free-space-to-chip atom pipeline. Numerical simulation verified that the reflection of the dipole laser is suppressed and that the atoms can be brought to the chip surface with a distance of only $100\,\mathrm{nm}$. Therefore, the pipeline allows a smooth transport of atoms from free space to the evanescent field trap of waveguides and promises a reliable atom source for a hybrid photonic-atom chip.
\end{abstract}
\maketitle

\section{Introduction}

In recent decades, neutral atoms have become one of the most important systems for realizing quantum information processing, and their potential applications have attracted increasing attention~\cite{monroe2002quantum,wehner2018quantum,Chang2018}. Quantum systems based on single atoms can demonstrate nonclassical effects and verify quantum theory~\cite{Beugnon2006}, and have been extended to quantum computing and quantum communication~\cite{saffman2010quantum,schafer2020tools,Weiss2017,briegel2000quantum,Specht2011}. Among these schemes, optical dipole traps have become a common tool for trapping and manipulating neutral cold atoms, which have a low atomic scattering rate, especially for far detuned frequencies~\cite{Schlosser2001,kaufman2021quantum}. Various configurations of optical dipole traps have been proposed and investigated in recent decades~\cite{Kim2019NC,Nayak19,liu2022multigrating,Frese00,ton2022state}. Additionally, optical cavities have been introduced to enhance the interaction between atoms and photons~\cite{Raimond2001,Reiserer2015}, and the strong coupling between arrays of atoms~\cite{Liu2022SXU,Chen2022,Deist2022} and multiple cavities~\cite{Daiss2021,Reiserer2022} are demonstrated recently. With these advantages, optical dipole traps and optical cavities have become a main platform for realizing atom-based quantum optics devices.

With the rapid development and maturity of integrated optical technology, quantum information devices call for integrating atoms into photonic chips~\cite{Chang2018,chang2019microring,beguin2020advanced,Luan2020tweezers,liu2022proposal,Wang2022chip}. Compared with conventional atom and cavity systems, such a hybrid photonic-atom chip system has many advantages, including the high stability and robustness of the system, strongly enhanced light-matter interaction with tightly confined optical modes, flexibility to engineer the long-range interaction, and great extensibility of devices on a single chip. In addition, it also has the potential to realize atom cooling by photonic chips~\cite{PrAchenliang,Dyer2022,Isichenko2022}. Therefore, the hybrid photonic-atom chip allows the cooling, transport, stable on-chip trapping, manipulation of single atoms, and promises a compact solution for quantum optics devices.

However, it is very challenging to bring cold atoms from the free space magneto-optics trap to the surface of the chip with a subwavelength distance~\cite{Kim2019NC,Zhou2023,Bouscal2023}. This is because the evanescent field on a photonic chip decays significantly in vacuum, thus a stable on-chip atom trapping demands an atom-surface distance of around $100\,\mathrm{nm}$, nonetheless the scattering of the dipole laser on the chip surface prevents the atoms from approaching the chip. While several methods have been proposed to transport cold atoms from free space to photonic chips~\cite{kuhr2001deterministic,thompson2013coupling,burgers2019clocked,Will2021}, achieving efficient and precise atom delivery beyond the limit of the "last one micron" remains a critical area of exploration for the advancement of atomic chips. One potential approach to circumvent the problem is introducing anti-reflection coating and the Brewster angle~\cite{jannasch2012nanonewton,cheng2016optimal,hoenig1991} to the photonics chip, which can prevent the scattering of light on the surface.

In this work, a hybrid photonic-atom chip platform is proposed for realizing an efficient free-space-to-chip pipeline, which allows efficient transport of single atoms to an on-chip evanescent field trap. A free space optical conveyor belt could be realized by the interference of the diffracted beam from the integrated apodized grating (AG) and the Gaussian beam, which intersects with a waveguide-based on-chip optical conveyor belt. By utilizing the anti-reflection of the Brewster angle, the reflection of the free space dipole trap on the waveguide surface is efficiently suppressed, which makes the atom delivery continuously from the free space to the integrated waveguide surface with a distance of around $100\,\mathrm{nm}$. The pipeline provides a potential solution for reliable and efficient single-atom sources on the chip, which is essential for future atom-based quantum optics devices.

The paper is organized as follows. Section II provides an overview of the operation procedure for the hybrid photonic-atom chip architecture. Section III covers the structure design of the atom pipeline and the impact of reflection. In Sec. IV, the reflection on the dielectric surface is analyzed, and it can be suppressed by the Brewster angle. Finally, Sec. V demonstrates the performance of the atom pipeline with the anti-reflection based on the Brewster angle.

\section{Overview}

\begin{figure}
\includegraphics[width=1\columnwidth]{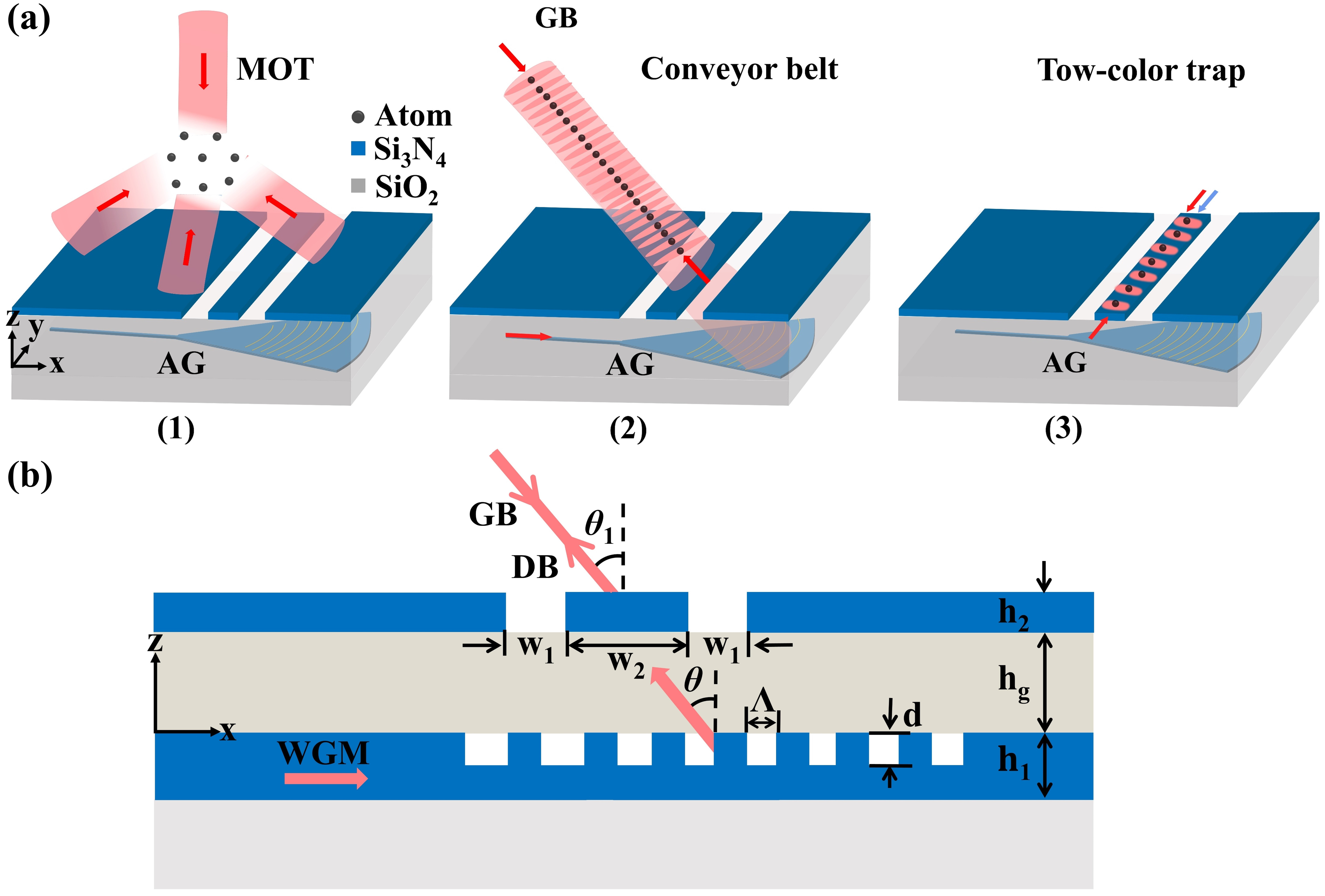}
\caption{(a) Flow chart illustration of the operation procedure for the hybrid photonic-atom chip architecture: (1) the preparation of a cold atom ensemble via a magneto-optics trap, (2) the free space to the chip transportation of atoms, and (3) the trapping and conveying of single atoms by the on-chip waveguides. (b) The cross-sectional geometry of the atom pipeline. GB: Gaussian beam, AG: apodized grating, WGM: waveguide mode, DB: diffraction beam.}
\label{fig1}
\end{figure}

Figure~\ref{fig1} sketches the two-layer chip architecture for realizing free-space-to-chip atom transportation. On the top layer, the photonic structures are fabricated to be vacuum-cladded, thus cold atoms can be transported and trapped on top of them. As demonstrated in previous works, the blue- and red-detuned lasers in the waveguide could create a dipole trap by evanescent fields with a trap depth of $\sim1\,$mK~\cite{Barnett2000,Hammes2003,le2004atoms,stievater2016modal}. The integrated waveguides and microring resonators not only allow the trap potential for manipulating the matter-wave of single atoms, but also enhance the photon-atom interaction due to the strongly confined optical fields on the chip, thus this platform provides a unique platform for realizing hybrid photonic-atomic quantum circuits~\cite{chang2019microring,Luan2020tweezers,Beguin2020,liu2022proposal,Wang2022chip}. The bottom layer of the chip provides the key optical components for converting the guided laser light in the waveguide into free space beams, preparing cold atoms and transporting the atoms to the top layer. The advantage of these on-chip components is a unified platform of free space optical components, allowing very compact realization of the whole system, which is very robust against perturbations and reduces the difficulties in aligning lasers. 

The working procedure of this architecture is as follows: 

(i) Preparation of a cold atom cloud with a distance of around $0.1-1\,\mathrm{mm}$ above the chip,  as shown in Fig.~\ref{fig1}(a)-(1). In most cold atom experiments, the cold atom ensemble can be efficiently prepared by a standard magneto-optics trap (MOT)~\cite{Metcalf1999}. We can choose a photonic chip with transparent substrates~\cite{liu2022proposal}, and thus a conventional six-beam MOT configuration could be realized, with the MOT center very close to the chip~\cite{Kim2019NC,Xulei2023}. 
It also has the potential to realize MOTs with on-chip optical devices. For instance, bottom-layer photonic circuits could provide three cooling laser beams via gratings, and the MOT could be realized by introducing an additional off-chip free space cooling laser and a printed circuit for the magnetic field gradient~\cite{PrAchenliang,Dyer2022,Isichenko2022}. 

(ii) Use an optical conveyor belt to transport cold atoms to the surface of the chip. An apodized grating (AG) on the bottom-layer could generate a focused Gaussian laser beam. The waveguide mode (WGM) on the under layer is diffracted by the apodized grating with a diffraction angle $\theta$ in the substrate and then output with a refraction angle $\theta_{1}$ in the air at wavelength $\lambda$ as shown in Fig.~\ref{fig1}(a)-(2). A Gaussian beam (GB) with the same wavelength $\lambda$ propagates in the opposite direction to interfere with the diffraction beam (DB) and forms an optical lattice above the surface of the chip, which enables the transport of atoms through the atomic conveyor belt technique~\cite{Kuhr2003}. By intersecting the free space standing-wave conveyor belt with the on-chip evanescent field-based conveyor belt~\cite{Schneeweiss2013}, atoms could be transported between the two dipole traps~\cite{Miroshnychenko2006} and an atom pipeline could be realized.

(iii) Trapping and guiding single atoms by the top-layer waveguide structures as shown in Fig.~\ref{fig1}(a)-(3). When the trapped atoms reach the chip surface (top layer) with a distance of about $100\,\mathrm{nm}$, the Casimir-Polder interaction increases rapidly~\cite{mclachlan1964van}, which will attach the atom to the chip surface and keep the atoms from moving to other positions on the chip. The blue-detuned TM (transverse magnetic) mode and the red-detuned TE (transverse electric) mode are incident on the waveguide. The evanescent field of the blue-detuned TM mode can provide a repulsive force to overcome the Casimir-Polder interaction and prevent the atom from being attached to the chip surface. The evanescent field of the red-detuned TE mode can provide an attractive force on the atom, and the combination of the evanescent field of the waveguide modes and the Casimir-Polder interaction forms the optical trap well, and its center is about $100\,\mathrm{nm}$ above the waveguide surface.

Following the above procedure, it is a potential method to transport the atoms from the MOT to the integrated photonic structures. However, as mentioned above, due to the potential diffraction and reflection of optical lasers around the surface, it is very challenging to realize a smooth intersection between the optical conveyor belt and the on-chip evanescent field traps. In this work, we focus on such an intersection and provide a practical solution of the pipeline connecting the free space and the chip. We consider the $^{87}$Rb atoms, with a D2 transition wavelength of $780\,\mathrm{nm}$. For the optical conveyor belt, we select a red-detuned $850\,\mathrm{nm}$ laser to form the standing-wave optical trap well. For the evanescent field trap, we select the blue-detuned TM mode with wavelength $\lambda_{b}=760\,\mathrm{nm}$ and the red-detuned TE mode with wavelength $\lambda_{r}=852\,\mathrm{nm}$ to propagate in the waveguide unidirectionally and bidirectionally, respectively. 

\section{Structure design for the pipeline}

The detailed design parameters for realizing the pipeline are illustrated in Fig.~\ref{fig1}(b). For the chip architecture, we choose $\mathrm{Si_{3}N_{4}}$ for the two layers of photonic structures, with thicknesses $h_1=300\,\mathrm{nm}$ and $h_2=240\,\mathrm{nm}$. It is worth noting that the $\lambda /4$ condition is ultilized as the anti-reflection condition for the $850\,\mathrm{nm}$ laser in the top $\mathrm{Si_{3}N_{4}}$ layer. Since the optical path length of light in the $\mathrm{Si_{3}N_{4}}$ layer is related to the incident angle, the thickness of the top $\mathrm{Si_{3}N_{4}}$ layer for anti-reflection is optimized by the numerical simulations. The bottom layer is buried in $\mathrm{SiO_2}$, with a distance from the top layers of $h_g=5\,\mathrm{\mu m}$. The apodized grating is etched on a tapered waveguide with a taper angle of $50^{\circ}$ and the apex of the taper up-surface is selected as the origin of the coordinate system, and the etching depth $d$ of the grating is set to be $200\,\mathrm{nm}$. The length of the grating part is about  $8\,\mathrm{\mu m}$, and the width of the input waveguide is $0.55\,\mathrm{\mu m}$. More details about the design of the apodized grating can be found in Ref.~\cite{liu2022multigrating}. The diffraction angle of the apodized grating is related to the grating period $\Lambda$. To form a Gaussian-like beam, the grating period $\Lambda$ of the apodized grating is gradually changed from $365\,\mathrm{nm}$ to $332\,\mathrm{nm}$, and the duty cycle $\eta$ increases from 0.5 to 0.3 along the grating. On the top layer, a waveguide intersects with the output Gaussian-like beam for transporting the atoms from the optical conveyor belt to the waveguide evanescent field trap. Here, to suppress the diffraction of the free space beam, the top layer of $\mathrm{Si_{3}N_{4}}$ is almost uniform, with only two grooves (width $w_1$) etched to form a waveguide (width $w_2$) in the center.

\begin{figure}
\includegraphics[width=1\columnwidth]{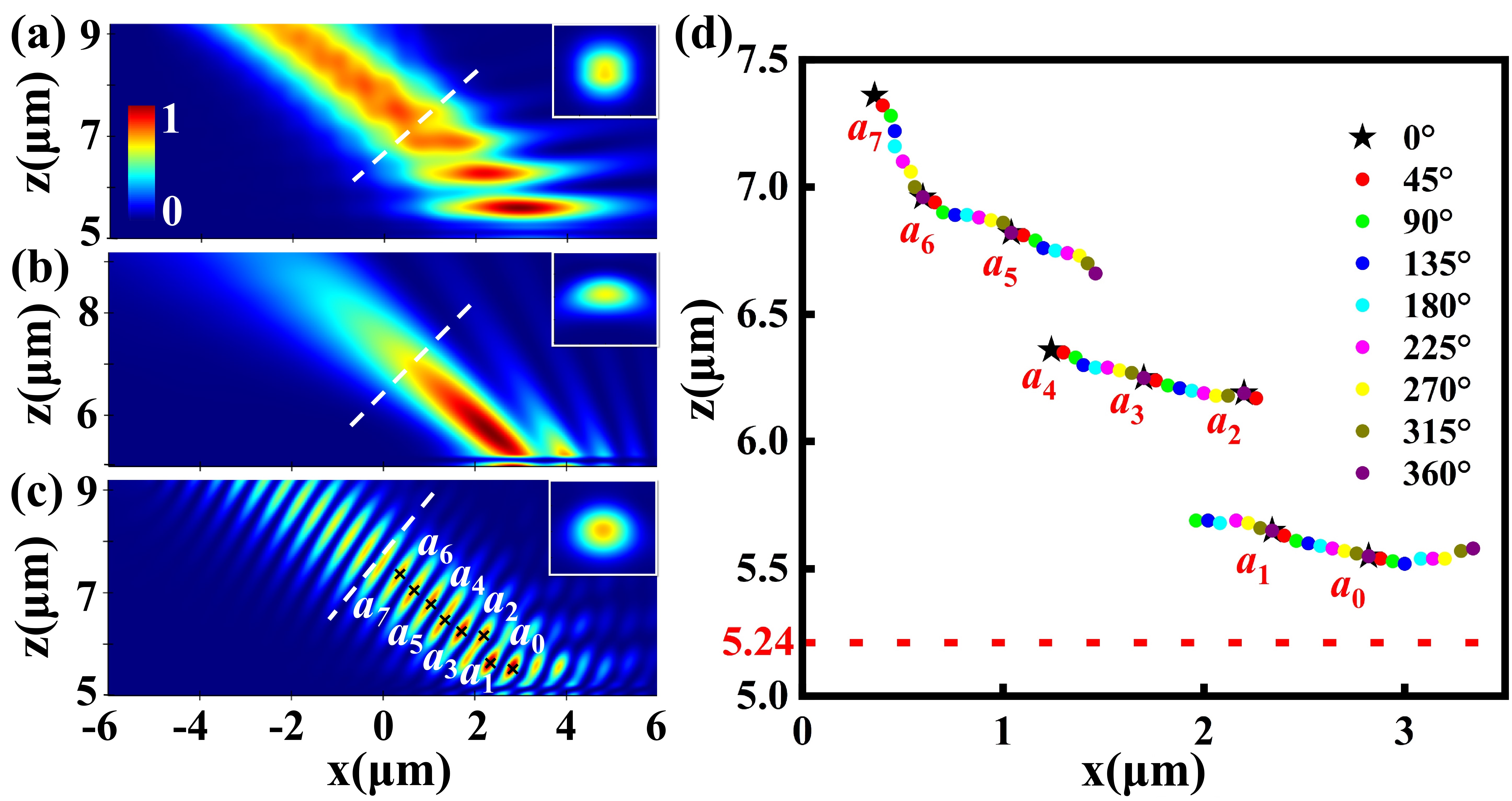}
\caption{The electric field distribution on the $x-z$ plane above the platform without etching grooves. (a) The intensity distribution of the Gaussian beam. (b) The intensity distribution of the diffraction beam from the integrated apodized grating. (c) The intensity distribution of the optical lattice. The insets in (a), (b) and (c) are the intensity distributions on the cross section denoted by the dotted lines. (d) The antinode position of the standing wave for different incident phase differences $\varDelta \varphi$.}
\label{fig2}
\end{figure}

Through three-dimensional numerical simulation of the electromagnetic field distribution by the finite-different time-domain (FDTD) method, the performance of the pipeline is investigated in Fig.~\ref{fig2}. 
To analyze the effect of the reflection on the chip surface, the top $\mathrm{Si_{3}N_{4}}$ layer without the etched grooves is considered first. A Gaussian beam of TE mode with a waist diameter of $3\,\mathrm{\mu m}$ is incident on the chip surface. At the same time, a counter-propagating beam from the apodized grating on the bottom-layer, which is the diffraction output from the TE mode, transmits through the top-layer with an angle of 51.525$^{\circ}$. To form the standing-wave fields for the optical conveyor belt, the Gaussian beam incident with the same angle of 51.525$^{\circ}$, which is different from the Brewster angle. The electric field distributions of the Gaussian beam and the diffraction beam on the $x-z$ plane above the chip is shown in Fig.~\ref{fig2}(a) and (b), respectively. Although the anti-reflection condition is taken into our design, we can still find non-negligible reflection of optical fields, which induced the interference field distribution close to the surface of the chip. As shown by the insets of Figs.~\ref{fig2}(a) and (b), the field distribution of the cross section shows a similar Gaussian-like beam profile. With input from both free space and the bottom-layer, the interference of the Gaussian beam gives rise to the standing-wave optical lattice, as shown in Fig.~\ref{fig2}(c). The antinodes of the standing wave are denoted by $a_{0}$, $a_{1}$, $a_{2}$,..., $a_{i}$, where the antinode with increasing index $i$ is far away from the chip surface.

For the wavelength of 850 nm, which is the red-detuned light for the $^{87}$Rb atom, the optical lattice can trap cold $^{87}$Rb atoms at the antinodes. Consequently, by controlling the phase difference $\varDelta \varphi$ between the two inputs for the Gaussian beam and the diffraction beam, it is anticipated that the atoms trapped in the antinodes could move toward the chip~\cite{Kuhr2003,Miroshnychenko2006,Kim2019NC}. In practical experiments, the two inputs are generated from the same laser, with phase control realized by an acousto-optics modulator and a radio-frequency (RF) signal generator~\cite{Xulei2023}. 
In Fig.~\ref{fig2}(d), the evolution of the antinode locations from $a_{0}$ to $a_{7}$ are plotted for various $\varDelta \varphi$, and the positions of $a_{i}$ with $\varDelta \varphi=0$ are denoted by stars.
If the antinode $a_{i}$ with $\varDelta \varphi=2\pi$ reaches the position of the antinode $a_{i-1}$ with $\varDelta \varphi=0$, the trapped atom can be transported from $a_{i}$ to $a_{i-1}$ by changing the phase difference $\varDelta \varphi$ of $2\pi$. 
For a desired atom pipeline, the antinode $a_{i}$ moves and connects with the antinode $a_{i-1}$ with a phase difference variation of $2\pi$, and atoms in the optical lattice can be continuously transported to $a_0$. However, we found that the traces of the antinodes are not continuous, and are separated into three sections. This means that the atoms from the free space, such as initially at $a_7$, could only be transported to about $1.5\,\mathrm{\mu m}$ above the chip and could not be transported to $a_2$ or $a_0$. The reason for this discontinuity is due to the reflection of the top-layer that forms an additional standing-wave pattern [Fig.~\ref{fig2}(a)] and prevents atom transport toward the chip surface.

\section{The suppression of reflection}

\begin{figure}
\includegraphics[width=1\columnwidth]{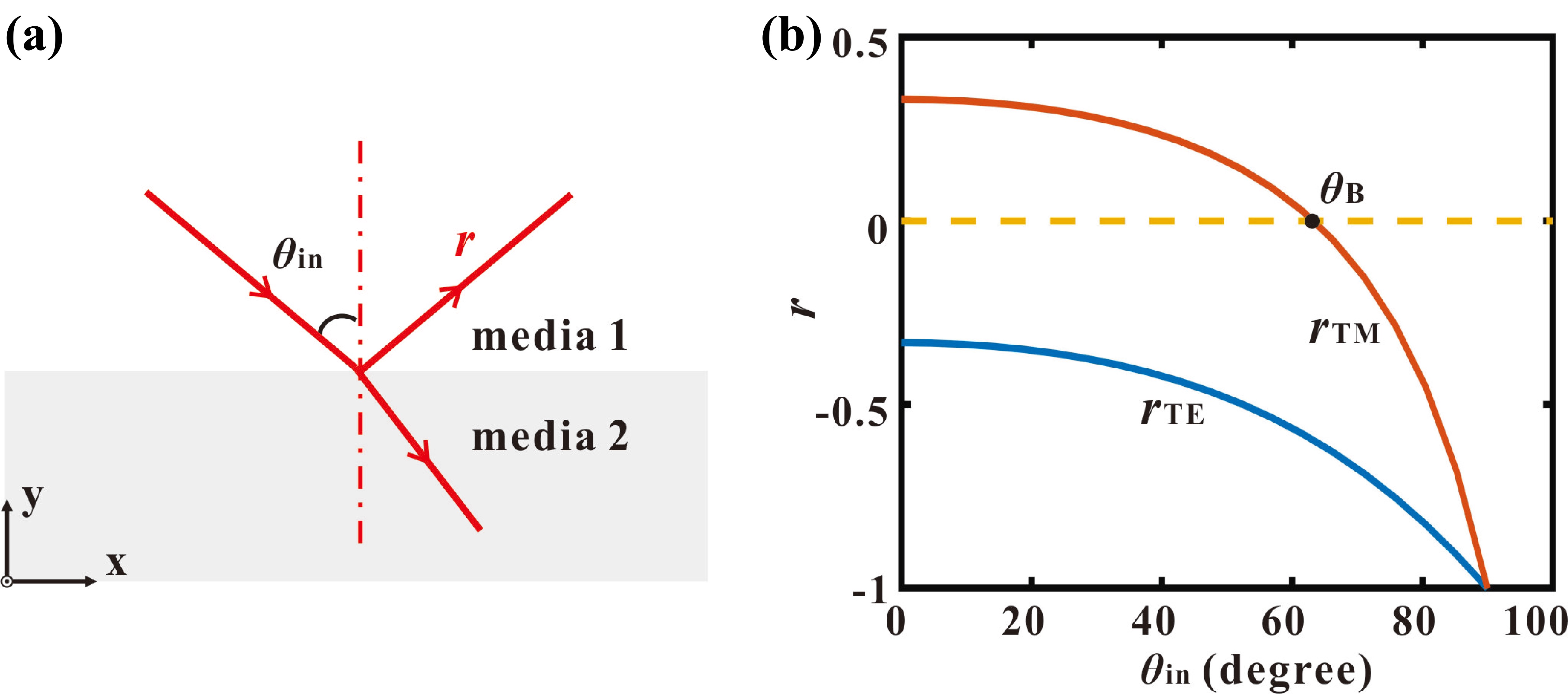}
\caption{The reflection and refraction of light at the interface of two media. (a) Illustration of light on the interface. (b) The reflection coefficients of TM and TE lights at $850\,\mathrm{nm}$  vary as the function of the incident angle $\theta_{in}$ on the air to Si3N4 interface.}
\label{fig3}
\end{figure}

The above results confirmed that a $~1\,\mathrm{\mu m}$-distance limits the application of optical conveyor belts in the hybrid photonic-atom chip, and there is considerable reflection of laser power even when the top-layer thickness satisfies the antireflection condition. To circumvent this problem, we introduce the  Brewster angle to further suppress the reflection, i.e. the Gaussian beam with the TM mode is incident on the chip surface with an input angle $\theta_{1}$ equal to the  Brewster angle of the top-layer. As shown in Fig.~\ref{fig3}(a), the reflection coefficients of optical fields on a flat surface for two polarizations follow~\cite{ra1973reflection}
\begin{eqnarray}
    r_{\mathrm{TM}} & = & \frac{n_{1}cos\theta_{in}-\sqrt{n_{2}^{2}-n_{1}^{2}sin^{2}\theta_{in}}}{n_{1}cos\theta_{in}+\sqrt{n_{2}^{2}-n_{1}^{2}sin^{2}\theta_{in}}},\label{eq1}\\
  r_{\mathrm{TE}} & = & \frac{n_{2}cos\theta_{in}-\sqrt{n_{1}^{2}-n_{1}^{4}sin^{2}\theta_{in}/n_{2}^{2}}}{n_{2}cos\theta_{in}+\sqrt{n_{1}^{2}-n_{1}^{4}sin^{2}\theta_{in}/n_{2}^{2}}},\label{eq2}
\end{eqnarray}
where $n_{1}$ and $n_{2}$ are the refractive index of media 1 to media 2. For our case, light with wavelength $850\,\mathrm{nm}$ incident from air ($n_{1}=1$) to $\mathrm{Si_{3}N_{4}}$ ($n_{2}$=1.999), the reflectivities $r_{\mathrm{TM}}$ and $r_{\mathrm{TE}}$ as functions of the incident angle $\theta_{in}$ are plotted in Fig.~\ref{fig2}(b). We found that at the so-called Brewster angle $\theta_{B}=63.4^{\circ}$, the reflection coefficients $r_{\mathrm{TM}}$ vanish, and the reflection is significantly suppressed. 

\begin{figure}
\includegraphics[width=1\columnwidth]{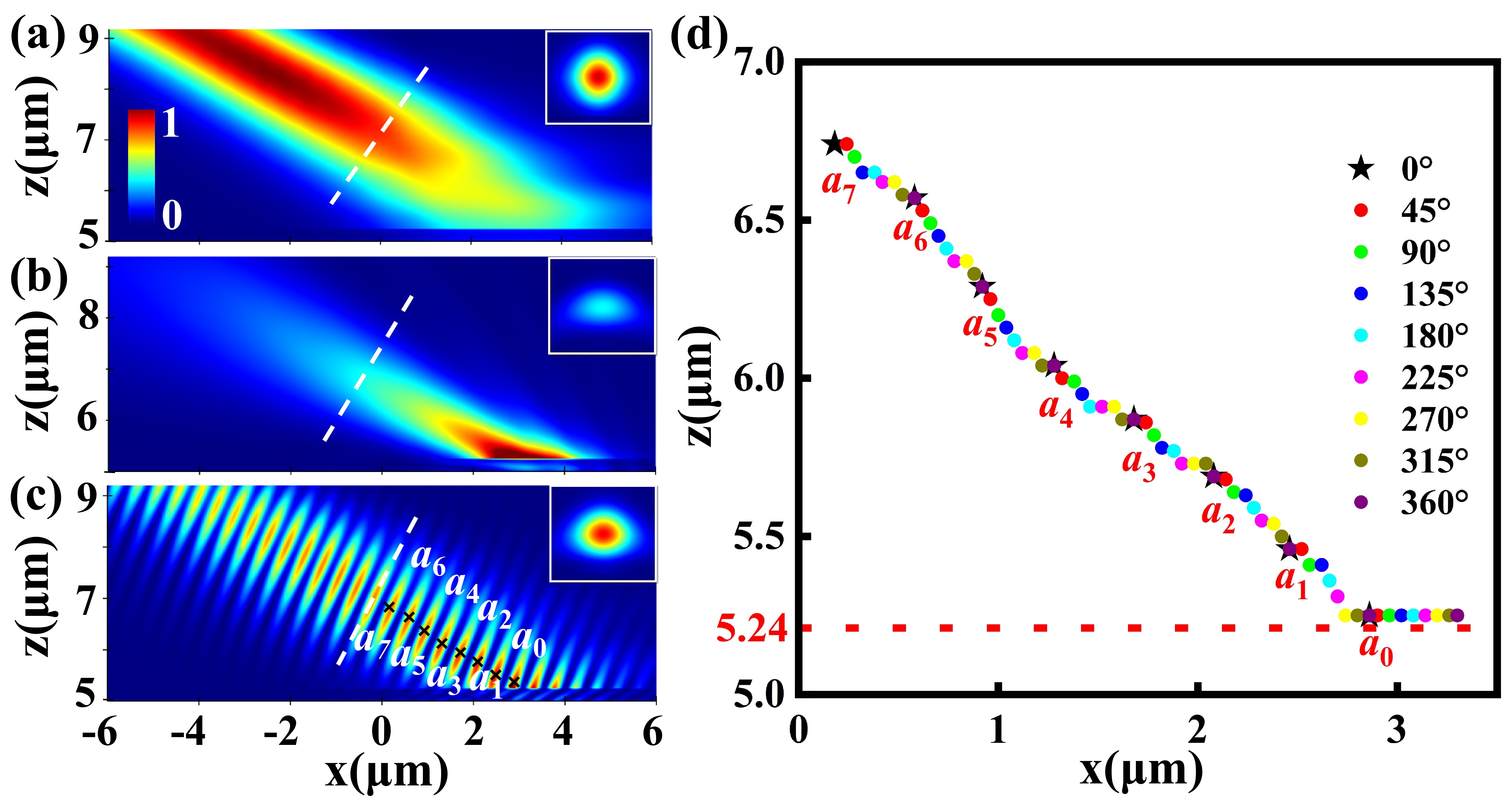}
\caption{The electric field distribution on the $x-z$ plane above the platform without etching grooves. (a) The intensity distribution of the Gaussian beam incident with the Brewster angle from free space. (b) The intensity distribution of the diffraction beam output with the Brewster angle from the integrated apodized grating. (c) The standing wave formed by the interference of the diffraction beam and the Gaussian beam. (d) The antinode position of the standing wave for different incident phase differences $\varDelta \varphi$.}
\label{fig4}
\end{figure}

According to Eq.~\ref{eq1}, the reflection of the Gaussian beam with TM mode will vanish when incident on the chip surface with the Brewster angle $\theta_{B}$. Compared with Fig.~\ref{fig2}(a), the reflection of the Gaussian beam on the chip surface, as shown in Fig.~\ref{fig4}(a), is suppressed both in the near field and the far field. The diffraction beam with the TM mode is output from the top layer with the same Brewster angle $\theta_{B}$, when the TM mode is incident on the apodized grating with grating periods $\Lambda$ ranging from $365\,\mathrm{nm}$ to $332\,\mathrm{nm}$. The intensity distribution of the diffraction beam on the $x-z$ plane is shown in Fig.~\ref{fig4}(b), with the inset as the intensity distribution on the cross section denoted by the white dotted line. The counter-propagating Gaussian beam and the diffraction beam interfere to form an optical lattice as shown in Fig.~\ref{fig4}(c). The relation between the antinode locations ($a_{0}$ to $a_{7}$) and the phase difference $\varDelta\varphi$ are given in Fig.~\ref{fig4}(d). In contrast to the discontinuous traces of antinodes, the scheme employing the Brewster angle generates a continuous trace of antinodes toward the interface, as all the antinodes $a_{i}$ with $\varDelta \varphi=2\pi$ move and connect to $a_{i-1}$ with $\varDelta \varphi=0$. As a result, the trapped atoms can be transported from the free space to very close to the chip surface. After the atoms are delivered to reach the chip surface, they can be trapped and transported further by the optical trap well on the waveguide surface, and the pipeline can be realized.

\section{The performances of the atom pipeline}  

\begin{figure}
\includegraphics[width=1\columnwidth]{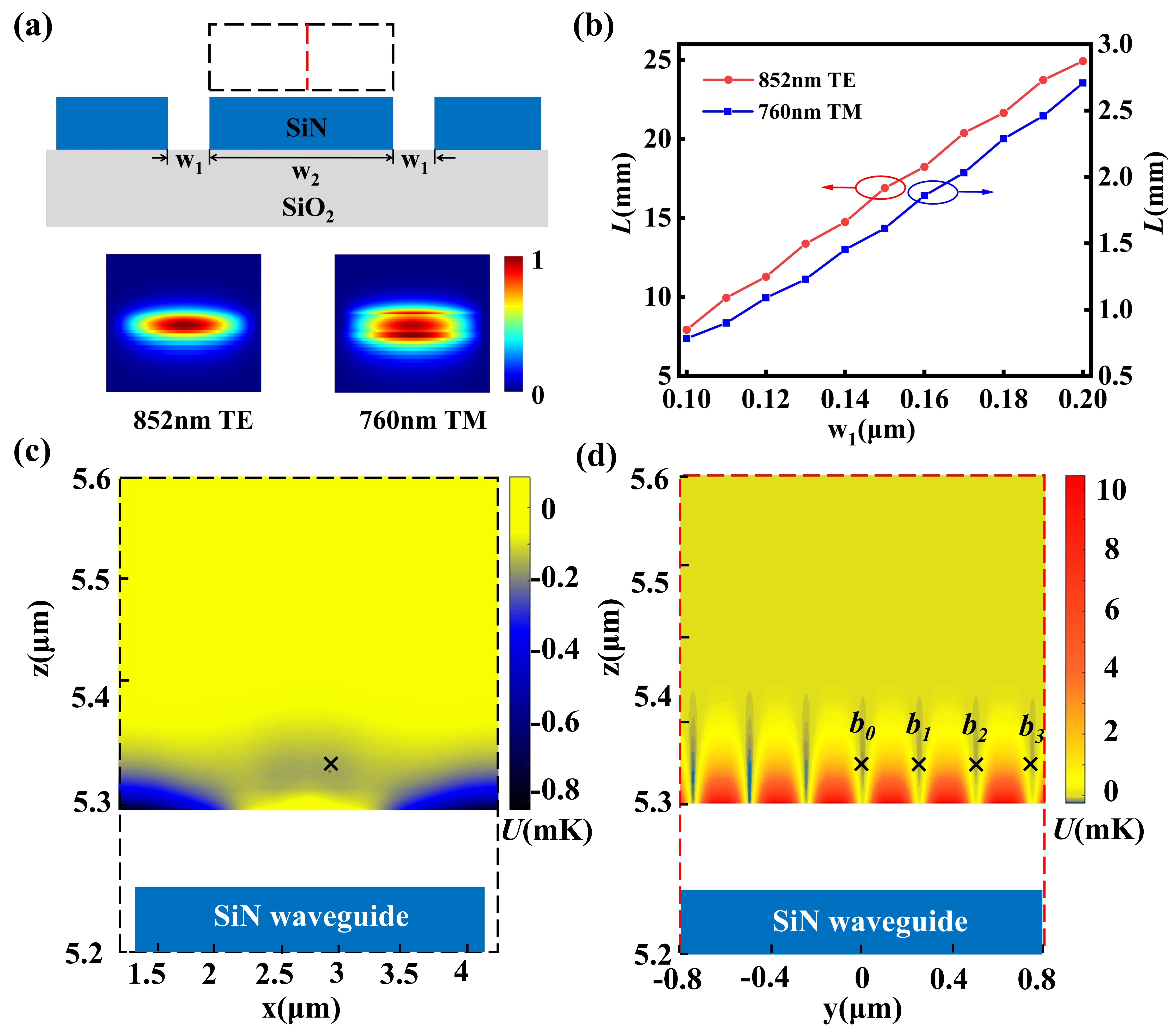}
\caption{The distribution of the optical trap depth. (a) The cross-section of the waveguide and the optical field distributions of the TE and TM modes. (b). The propagation lengths of the waveguide modes vary with the gap width. (c)-(d) Optical trap depth formed by the evanescent field of the waveguide modes and the Casimir-Polder interaction (c) on the $x-z$ plane and (d) on the $y-z$ plane.}
\label{fig5}
\end{figure}

In practice, we should fabricate waveguides for the smooth pipeline connecting the free space MOT and the on-chip evanescent field atom trap. On the top layer of the chip, two grooves are etched to form a waveguide, which intersects with the optical lattice at the right angle. On the one hand, we should make a waveguide with a small cross-section to effectively enhance the electric field intensity of the waveguide mode, thus saving the required laser power for the evanescent field trap. On the other hand, the width of the two grooves should be as small as possible because the etched grooves may influence on the optical lattice by introducing the scattering of the dipole trap laser. Considering such a trade-off relation, we choose $w_1=0.15\,\mathrm{\mu m}$ and $w_2=3\,\mathrm{\mu m}$. Figure~\ref{fig5}(a) depicts the cross-section structure of the waveguide,  plots the corresponding electric-field distribution for blue- and red-detuned trap lasers, and shows confined modes in the waveguide region. Due to the leakage loss to the unetached slab for a finite groove width, the mode intensity should decay with propagation as $\propto \mathrm{exp}(-z/L)$, with $L$ being the mode propagation length in the waveguide. Figure~\ref{fig5}(b) numerically investigated the propagation length against $w_1$, and both modes have a propagation length exceeding $1\,\mathrm{mm}$, which indicates an energy loss less than $1\%$ for a waveguide length less than $10\,\mathrm{\mu m}$ when realizing the pipeline. In the area of the chip surface covered by the free space dipole laser beam, which has a diameter of several micrometers, minimizing the grooves around the waveguide are minimized for a small diffraction loss of the beam, while the width of the waveguide is increased and the length of this narrow groove regime is $10\,\mathrm{\mu m}$ to suppress the leakage of waveguide modes into the unetched slab. Outside the laser spot area, a tapered waveguide can be used to narrow the waveguide, resulting in stronger coupling between the atom and the waveguide mode. Additionally, the width of the groove can be increased to avoid leakage loss. Here, the difference between the TE and TM modes is attributed to the different electric-field boundary conditions for orthogonal optical polarizations and thus different leakage losses through tunneling the gap $w_1$. 

In the following, the optical trap depth is analyzed. By ignoring the Zeeman sublevels, the optical dipole potential formed by the linear polarization light can be estimated as
\begin{equation}
U=\frac{\hbar\gamma I_{0}}{24I_{S}}(\frac{1}{\delta_{1/2}}+\frac{2}{\delta_{3/2}}),
\label{eq3}
\end{equation}
where $\gamma/2\pi \approx 6.1\,\mathrm{MHz}$ is the natural linewidth of the $^{87}$Rb D2 transition, $I_{S}$ is the saturation intensity the cyclic $\sigma^+$ transition, $I_{0}$ is the intensity of the optical field, $\delta_{1/2}$ and $\delta_{3/2}$ are the detunings between the light frequency and the $D_{1}$ and $D_{2}$ transitions, which exceed the hyperfine splitting of the excited state. For the $^{87}$Rb atom with a resonant wavelength of about $780\,\mathrm{nm}$, the gradient force formed by a red-detuned optical field ($850$ or $852\,\mathrm{nm}$) is an attractive force, and that formed by a blue-detuned optical field ($760\,\mathrm{nm}$) is a repulsive force. In the calculation of the optical dipole potential, the reference point is set at infinity with $U=0$. 

When the atom approaches the surface of chip with a small distance, the Casimir-Polder potential also comes into play, which can be estimated as $U_{CP}=-0.12\hbar \gamma \lambda^3/3\pi^3 d^3$ with $d$ represents the distance from the surface~\cite{mclachlan1964van}. The combination of the two-color evanescent field of the waveguide modes and the Casimir-Polder interaction results in an optical trap well above the surface of the waveguide. Figures~\ref{fig5}(c) and (d) show the optical trap depths on the $x-z$ plane and $x-y$ plane above the waveguide surface corresponding to the black dashed
rectangle and the red line in Fig.~\ref{fig5}(a), where the powers of the blue- and red-detuned modes are $100\,\mathrm{mW}$ and $77.7\,\mathrm{mW}$, respectively. A lattice of the trap well denoted by $b_{0}$, $b_{1}$,...$b_{i}$ along the waveguide is generated, as shown in Fig.~\ref{fig5}(d), with a distance between the trap centers and the waveguide surface of about $100\,\mathrm{nm}$. By manipulating the phase difference $\varDelta \phi$ of the bidirectionally input red-detuned TE modes, the trap wells will move along the waveguide, which will transport the trapped atom to other positions of the chip. As a tradeoff for the reduced perturbations of the waveguide structure to the free space beams, the relatively wide waveguide requires a higher laser power for near-field optical dipole trap. A high power up to 10 W has been experimentally demonstrated for a waveguide mode on the optical chips~\cite{zhang2023high}, which makes it realistic for the laser power required in our designed waveguide. Though the tightly confined optical modes in the waveguide could induces the out-of-phase longitudinal polarization component, which may have an impact on potential of the atoms trapped in the evanescent field of a waveguide~\cite{lodahl2017chiral,junge2013strong}. This impact on the potential is not discussed, since this manuscript mainly focuses on the transporting of atoms from free space to the chip surface.

\begin{figure}
\includegraphics[width=1\columnwidth]{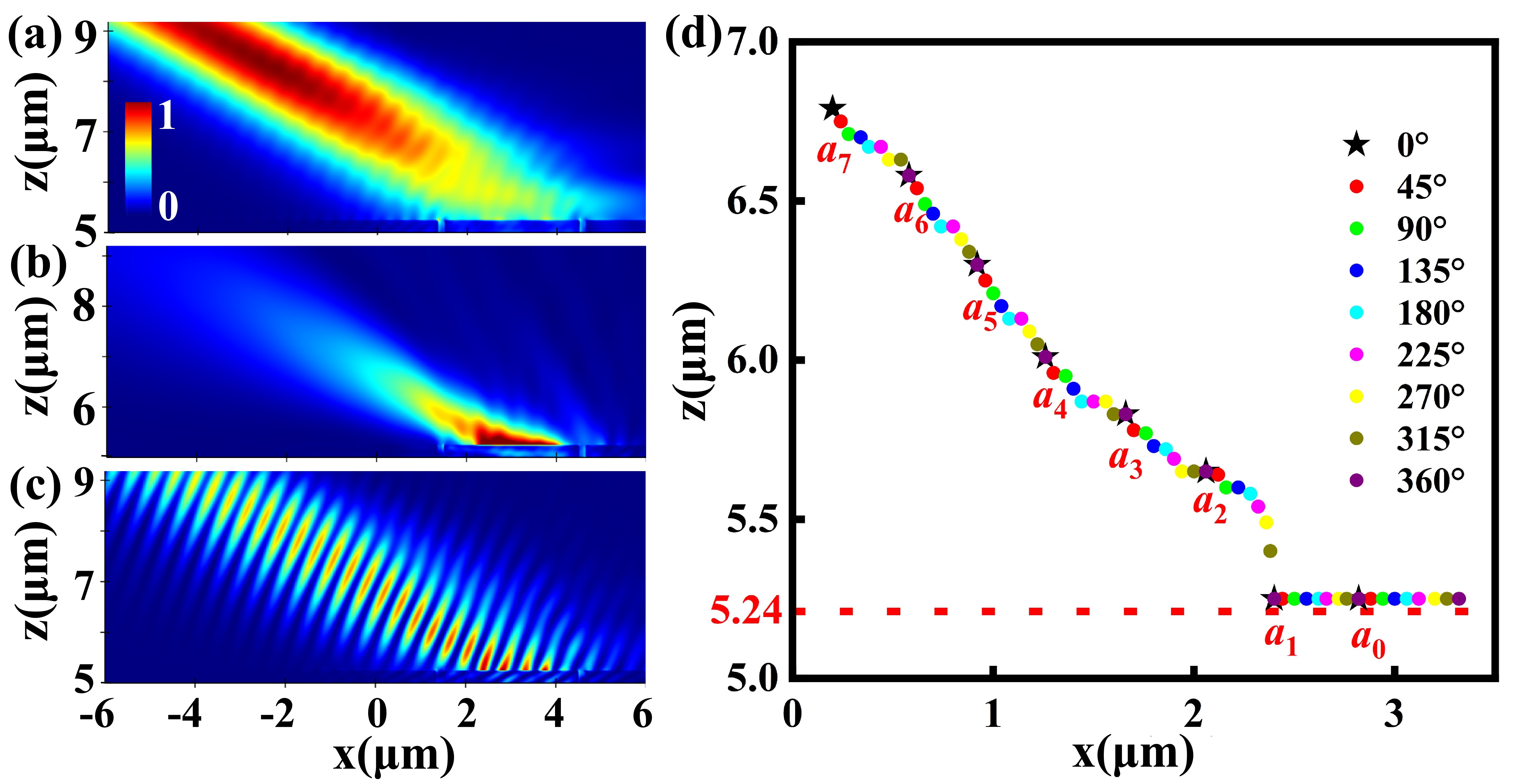}
\caption{The free space optical conveyor belt formed by the platform with etching grooves. (a)-(c) The intensity distributions of the optical potential in the $x-z$ plane. (d) The antinode positions against the phase differences $\varDelta \varphi$.}
\label{fig6}
\end{figure}

For the free space conveyor belt, the intensity distributions for the case with etched grooves are shown in Figs.~\ref{fig6}(a)-(c). Compared with Fig.~\ref{fig4}(a)-(c), the scattering from the grooves induces a minor perturbation on the optical lattice's intensity distribution, leading to a significant drop of the antinode position close to a1. When the atom approaches the waveguide surface to within hundreds of nanometer, the optical trap formed by the waveguide modes is employed to prevent the atom from attaching to the chip surface and to maintain a separation of about $100\,\mathrm{nm}$ for the atom-waveguide coupling. Therefore, the sharp drop caused by the etched grooves does not destroy the atom delivery from the free space to the near field of the waveguide surface.

\begin{figure}
\includegraphics[width=1\columnwidth]{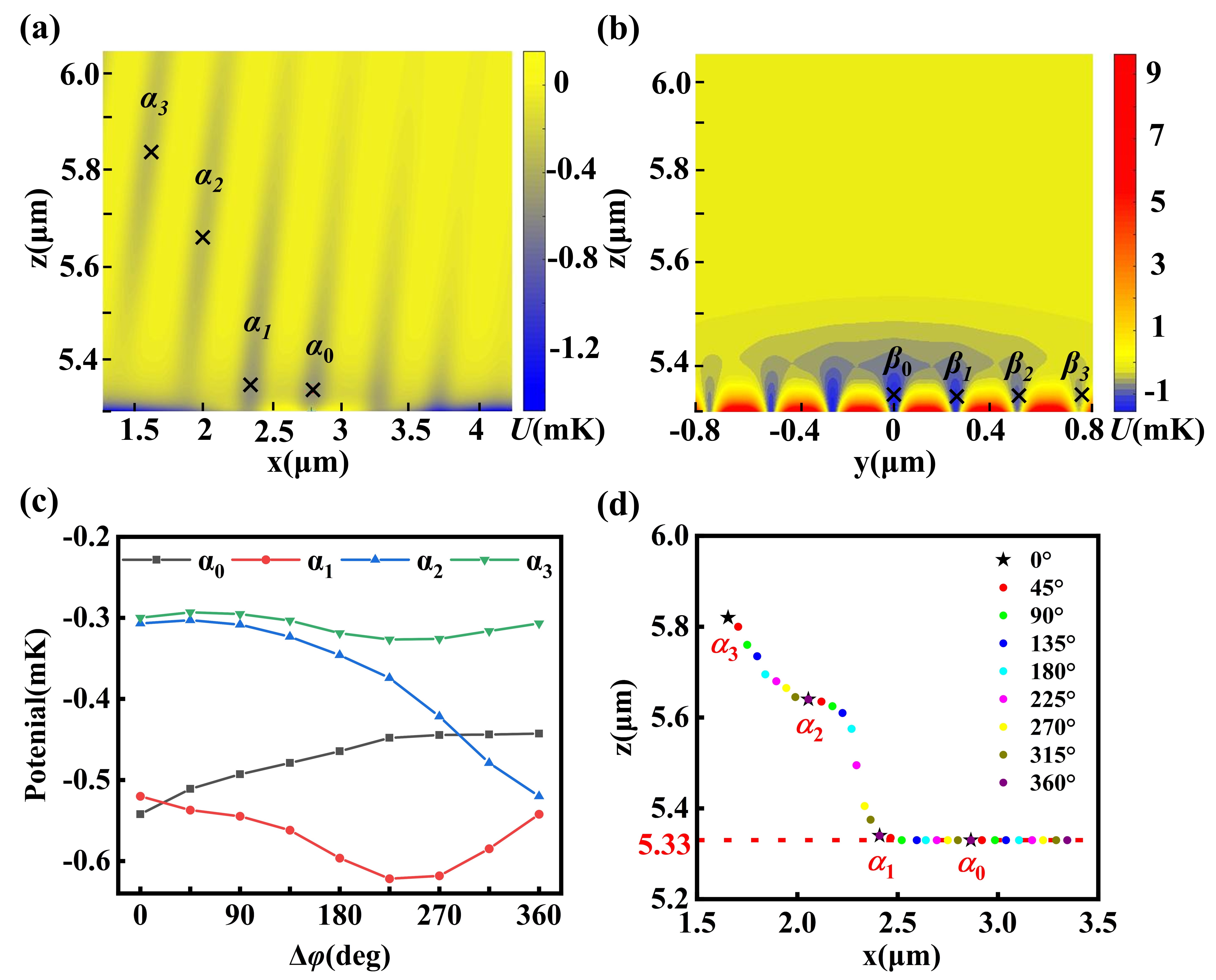}
\caption{The distribution of the total optical trap depth. (a) The total optical trap potential distribution on the $x-z$ plane. (b) The total optical trap depth potential distribution in the $x-y$ plane. (c) The effective optical trap depths of the trap wells $\alpha_{0}$ to $\alpha_{3}$ as functions of the phase difference $\varDelta \varphi$. (d) The positions of the trap centers against the phase difference $\varDelta \varphi$.}
\label{fig7}
\end{figure}

Last, when realizing the atom pipeline, the optical trap potential on the intersection is a superposition of potentials from the waveguide modes, the free space standing-wave, and the Casimir-Polder interaction. The distributions of the total optical trap potential of the intersection region with $\varDelta \phi=0$ and $\varDelta \varphi=0$ for both free space and on-chip optical fields on the $x-z$ plane are shown in Fig.~\ref{fig7}(a), and those on the $x-y$ plane are shown in Fig.~\ref{fig7}(b). The powers of the Gaussian beam and the incident waveguide mode are $1.69\,\mathrm{mW}$ and $100\,\mathrm{mW}$, respectively. The locations of trap wells are denoted by $\alpha_{i}$ ($i\in{0,1,2,...}$) along the free-space standing wave, and are denoted by $\beta_{\pm i}$ ($i\in{0,\pm1,\pm2,...}$) along the waveguide. By manipulating the phase difference $\varDelta \varphi$ of the free-space standing wave, the trap wells move and the effective trap depths vary. For example,  Fig.~\ref{fig7}(c) shows the results for $\alpha_{0}$ to $\alpha_{3}$ with the optical fields in the waveguide fixed ($\varDelta \phi=0$). To ensure a stable atom trap during atom delivery, trap depths deeper than $0.3\,\mathrm{mK}$ are maintained. The trap centers of the wells vary with the phase difference $\varDelta \varphi$, which shows a continue movement between the adjacent trap wells. Near the waveguide surface, trap well $\alpha_{0}$ connects with trap well $\beta_{0}$. The position of trap well $\beta_{0}$ will move horizontally to the adjacent trap well by manipulating the phase difference $\varDelta \phi$ of the two red detuned TM modes, which will transport the atoms along the waveguide surface. Although the atom delivery of about several micrometer above the chip surface is considered above, it can also be achieved further away from the chip surface where a stable standing wave is achievable without the influence of reflection. Then, a continuous free-space-to-chip pipeline for single atoms is realized on the integrated chip.

\section{Conclusion}

A two-layer photonic chip architecture is proposed for realizing a compact hybrid photonic-atom chip. In particular, a key ingredient for realizing the efficient transportation of cold atoms from the free space magneto-optics trap to the integrated waveguide evanescent field trap is proposed and numerically validated. A free space optical conveyor belt is formed by the interference of an on-chip diffracted laser beam and a free space Gaussian light. However, due to the diffraction and reflection of the dipole laser, the atoms could only be transported close to the chip surface with a distance of about one wavelength. Thus, a free-space-to-chip pipeline for smooth delivery of single atoms is realized by introducing the Brewster angle to eliminate the reflection of dipole lasers on the surface of the chip. The continuous delivery of cold atoms from free space to the chip surface with a distance less than $100\,\mathrm{nm}$ enables efficient interaction between the atoms and the evanescent field of the guided mode in the waveguides. Therefore, our proposal provides a reliable atom source for constructing on-chip hybrid atom-photonic devices, including single-photon sources, high-fidelity single-photon quantum gates, and quantum memories. Such a platform also promises a new type of integrated circuit for matter-waves of single atoms~\cite{Amico2022}. 

\section{Acknowledgments}
We would like to thank Lei Xu and Ling-Xiao Wang for helpful discussions. This work was funded by the National Key R\&D Program (Grant No.~2021YFA1402004 and No.~2018YFA0306400), the National Natural Science Foundation of China (Grants No.~U21A6006, No.~12104441, No.~92265210, No.~U21A20433, No.~62061160487 and No.~12074194), the Natural Science Foundation of Anhui Provincial (Grant No.~2108085MA17 and No.~2108085MA22), the Key Research and Development Program of Anhui Province (Grant No.~2022b1302007), and the Industrial Prospect and Key Core Technology Projects of Jiangsu Provincial Key R\&D Program (Grant No.~BE2022071). CLZ was also supported by the Fundamental Research Funds for the Central Universities and USTC Research Funds of the Double First-Class Initiative. The numerical calculations in this paper have been done on the supercomputing system in the Supercomputing  Center of the University of Science and Technology of China. This work was partially carried out at the USTC Center for Micro and Nanoscale Research and Fabrication. 

\bibliography{AtomPipeline}

\begin{thebibliography}{56}%
\makeatletter
\providecommand \@ifxundefined [1]{%
 \@ifx{#1\undefined}
}%
\providecommand \@ifnum [1]{%
 \ifnum #1\expandafter \@firstoftwo
 \else \expandafter \@secondoftwo
 \fi
}%
\providecommand \@ifx [1]{%
 \ifx #1\expandafter \@firstoftwo
 \else \expandafter \@secondoftwo
 \fi
}%
\providecommand \natexlab [1]{#1}%
\providecommand \enquote  [1]{``#1''}%
\providecommand \bibnamefont  [1]{#1}%
\providecommand \bibfnamefont [1]{#1}%
\providecommand \citenamefont [1]{#1}%
\providecommand \href@noop [0]{\@secondoftwo}%
\providecommand \href [0]{\begingroup \@sanitize@url \@href}%
\providecommand \@href[1]{\@@startlink{#1}\@@href}%
\providecommand \@@href[1]{\endgroup#1\@@endlink}%
\providecommand \@sanitize@url [0]{\catcode `\\12\catcode `\$12\catcode
  `\&12\catcode `\#12\catcode `\^12\catcode `\_12\catcode `\%12\relax}%
\providecommand \@@startlink[1]{}%
\providecommand \@@endlink[0]{}%
\providecommand \url  [0]{\begingroup\@sanitize@url \@url }%
\providecommand \@url [1]{\endgroup\@href {#1}{\urlprefix }}%
\providecommand \urlprefix  [0]{URL }%
\providecommand \Eprint [0]{\href }%
\providecommand \doibase [0]{http://dx.doi.org/}%
\providecommand \selectlanguage [0]{\@gobble}%
\providecommand \bibinfo  [0]{\@secondoftwo}%
\providecommand \bibfield  [0]{\@secondoftwo}%
\providecommand \translation [1]{[#1]}%
\providecommand \BibitemOpen [0]{}%
\providecommand \bibitemStop [0]{}%
\providecommand \bibitemNoStop [0]{.\EOS\space}%
\providecommand \EOS [0]{\spacefactor3000\relax}%
\providecommand \BibitemShut  [1]{\csname bibitem#1\endcsname}%
\let\auto@bib@innerbib\@empty
\bibitem [{\citenamefont {Monroe}(2002)}]{monroe2002quantum}%
  \BibitemOpen
  \bibfield  {author} {\bibinfo {author} {\bibfnamefont {C.}~\bibnamefont
  {Monroe}},\ }\href {\doibase 10.1038/416238a} {\bibfield  {journal} {\bibinfo
   {journal} {Nature}\ }\textbf {\bibinfo {volume} {416}},\ \bibinfo {pages}
  {238} (\bibinfo {year} {2002})}\BibitemShut {NoStop}%
\bibitem [{\citenamefont {Wehner}\ \emph {et~al.}(2018)\citenamefont {Wehner},
  \citenamefont {Elkouss},\ and\ \citenamefont {Hanson}}]{wehner2018quantum}%
  \BibitemOpen
  \bibfield  {author} {\bibinfo {author} {\bibfnamefont {S.}~\bibnamefont
  {Wehner}}, \bibinfo {author} {\bibfnamefont {D.}~\bibnamefont {Elkouss}}, \
  and\ \bibinfo {author} {\bibfnamefont {R.}~\bibnamefont {Hanson}},\ }\href
  {\doibase 10.1126/science.aam928} {\bibfield  {journal} {\bibinfo  {journal}
  {Science}\ }\textbf {\bibinfo {volume} {362}},\ \bibinfo {pages} {eaam9288}
  (\bibinfo {year} {2018})}\BibitemShut {NoStop}%
\bibitem [{\citenamefont {Chang}\ \emph {et~al.}(2018)\citenamefont {Chang},
  \citenamefont {Douglas}, \citenamefont {Gonz{\'{a}}lez-Tudela}, \citenamefont
  {Hung},\ and\ \citenamefont {Kimble}}]{Chang2018}%
  \BibitemOpen
  \bibfield  {author} {\bibinfo {author} {\bibfnamefont {D.~E.}\ \bibnamefont
  {Chang}}, \bibinfo {author} {\bibfnamefont {J.~S.}\ \bibnamefont {Douglas}},
  \bibinfo {author} {\bibfnamefont {A.}~\bibnamefont {Gonz{\'{a}}lez-Tudela}},
  \bibinfo {author} {\bibfnamefont {C.-L.}\ \bibnamefont {Hung}}, \ and\
  \bibinfo {author} {\bibfnamefont {H.~J.}\ \bibnamefont {Kimble}},\ }\href
  {\doibase 10.1103/RevModPhys.90.031002} {\bibfield  {journal} {\bibinfo
  {journal} {Reviews of Modern Physics}\ }\textbf {\bibinfo {volume} {90}},\
  \bibinfo {pages} {031002} (\bibinfo {year} {2018})}\BibitemShut {NoStop}%
\bibitem [{\citenamefont {Beugnon}\ \emph {et~al.}(2006)\citenamefont
  {Beugnon}, \citenamefont {Jones}, \citenamefont {Dingjan}, \citenamefont
  {Darqui{\'{e}}}, \citenamefont {Messin}, \citenamefont {Browaeys},\ and\
  \citenamefont {Grangier}}]{Beugnon2006}%
  \BibitemOpen
  \bibfield  {author} {\bibinfo {author} {\bibfnamefont {J.}~\bibnamefont
  {Beugnon}}, \bibinfo {author} {\bibfnamefont {M.~P.~a.}\ \bibnamefont
  {Jones}}, \bibinfo {author} {\bibfnamefont {J.}~\bibnamefont {Dingjan}},
  \bibinfo {author} {\bibfnamefont {B.}~\bibnamefont {Darqui{\'{e}}}}, \bibinfo
  {author} {\bibfnamefont {G.}~\bibnamefont {Messin}}, \bibinfo {author}
  {\bibfnamefont {A.}~\bibnamefont {Browaeys}}, \ and\ \bibinfo {author}
  {\bibfnamefont {P.}~\bibnamefont {Grangier}},\ }\href {\doibase
  10.1038/nature04628} {\bibfield  {journal} {\bibinfo  {journal} {Nature}\
  }\textbf {\bibinfo {volume} {440}},\ \bibinfo {pages} {779} (\bibinfo {year}
  {2006})},\ \Eprint {http://arxiv.org/abs/0610149} {arXiv:0610149 [quant-ph]}
  \BibitemShut {NoStop}%
\bibitem [{\citenamefont {Saffman}\ \emph {et~al.}(2010)\citenamefont
  {Saffman}, \citenamefont {Walker},\ and\ \citenamefont
  {M{\o}lmer}}]{saffman2010quantum}%
  \BibitemOpen
  \bibfield  {author} {\bibinfo {author} {\bibfnamefont {M.}~\bibnamefont
  {Saffman}}, \bibinfo {author} {\bibfnamefont {T.~G.}\ \bibnamefont {Walker}},
  \ and\ \bibinfo {author} {\bibfnamefont {K.}~\bibnamefont {M{\o}lmer}},\
  }\href {\doibase 10.1103/RevModPhys.82.2313} {\bibfield  {journal} {\bibinfo
  {journal} {Reviews of modern physics}\ }\textbf {\bibinfo {volume} {82}},\
  \bibinfo {pages} {2313} (\bibinfo {year} {2010})}\BibitemShut {NoStop}%
\bibitem [{\citenamefont {Sch{\"a}fer}\ \emph {et~al.}(2020)\citenamefont
  {Sch{\"a}fer}, \citenamefont {Fukuhara}, \citenamefont {Sugawa},
  \citenamefont {Takasu},\ and\ \citenamefont {Takahashi}}]{schafer2020tools}%
  \BibitemOpen
  \bibfield  {author} {\bibinfo {author} {\bibfnamefont {F.}~\bibnamefont
  {Sch{\"a}fer}}, \bibinfo {author} {\bibfnamefont {T.}~\bibnamefont
  {Fukuhara}}, \bibinfo {author} {\bibfnamefont {S.}~\bibnamefont {Sugawa}},
  \bibinfo {author} {\bibfnamefont {Y.}~\bibnamefont {Takasu}}, \ and\ \bibinfo
  {author} {\bibfnamefont {Y.}~\bibnamefont {Takahashi}},\ }\href {\doibase
  10.1038/s42254-020-0195-3} {\bibfield  {journal} {\bibinfo  {journal} {Nat.
  Rev. Phys.}\ }\textbf {\bibinfo {volume} {2}},\ \bibinfo {pages} {411}
  (\bibinfo {year} {2020})}\BibitemShut {NoStop}%
\bibitem [{\citenamefont {Weiss}\ and\ \citenamefont
  {Saffman}(2017)}]{Weiss2017}%
  \BibitemOpen
  \bibfield  {author} {\bibinfo {author} {\bibfnamefont {D.~S.}\ \bibnamefont
  {Weiss}}\ and\ \bibinfo {author} {\bibfnamefont {M.}~\bibnamefont
  {Saffman}},\ }\href {\doibase 10.1063/PT.3.3626} {\bibfield  {journal}
  {\bibinfo  {journal} {Physics Today}\ }\textbf {\bibinfo {volume} {70}},\
  \bibinfo {pages} {44} (\bibinfo {year} {2017})}\BibitemShut {NoStop}%
\bibitem [{\citenamefont {Briegel}\ \emph {et~al.}(2000)\citenamefont
  {Briegel}, \citenamefont {Calarco}, \citenamefont {Jaksch}, \citenamefont
  {Cirac},\ and\ \citenamefont {Zoller}}]{briegel2000quantum}%
  \BibitemOpen
  \bibfield  {author} {\bibinfo {author} {\bibfnamefont {H.-J.}\ \bibnamefont
  {Briegel}}, \bibinfo {author} {\bibfnamefont {T.}~\bibnamefont {Calarco}},
  \bibinfo {author} {\bibfnamefont {D.}~\bibnamefont {Jaksch}}, \bibinfo
  {author} {\bibfnamefont {J.~I.}\ \bibnamefont {Cirac}}, \ and\ \bibinfo
  {author} {\bibfnamefont {P.}~\bibnamefont {Zoller}},\ }\href {\doibase
  10.1080/09500340008244052} {\bibfield  {journal} {\bibinfo  {journal}
  {Journal of modern optics}\ }\textbf {\bibinfo {volume} {47}},\ \bibinfo
  {pages} {415} (\bibinfo {year} {2000})}\BibitemShut {NoStop}%
\bibitem [{\citenamefont {Specht}\ \emph {et~al.}(2011)\citenamefont {Specht},
  \citenamefont {N{\"{o}}lleke}, \citenamefont {Reiserer}, \citenamefont
  {Uphoff}, \citenamefont {Figueroa}, \citenamefont {Ritter},\ and\
  \citenamefont {Rempe}}]{Specht2011}%
  \BibitemOpen
  \bibfield  {author} {\bibinfo {author} {\bibfnamefont {H.~P.}\ \bibnamefont
  {Specht}}, \bibinfo {author} {\bibfnamefont {C.}~\bibnamefont
  {N{\"{o}}lleke}}, \bibinfo {author} {\bibfnamefont {A.}~\bibnamefont
  {Reiserer}}, \bibinfo {author} {\bibfnamefont {M.}~\bibnamefont {Uphoff}},
  \bibinfo {author} {\bibfnamefont {E.}~\bibnamefont {Figueroa}}, \bibinfo
  {author} {\bibfnamefont {S.}~\bibnamefont {Ritter}}, \ and\ \bibinfo {author}
  {\bibfnamefont {G.}~\bibnamefont {Rempe}},\ }\href {\doibase
  10.1038/nature09997} {\bibfield  {journal} {\bibinfo  {journal} {Nature}\
  }\textbf {\bibinfo {volume} {473}},\ \bibinfo {pages} {190} (\bibinfo {year}
  {2011})},\ \Eprint {http://arxiv.org/abs/1103.1528} {arXiv:1103.1528}
  \BibitemShut {NoStop}%
\bibitem [{\citenamefont {Schlosser}\ \emph {et~al.}(2001)\citenamefont
  {Schlosser}, \citenamefont {Reymond}, \citenamefont {Protsenko},\ and\
  \citenamefont {Grangier}}]{Schlosser2001}%
  \BibitemOpen
  \bibfield  {author} {\bibinfo {author} {\bibfnamefont {N.}~\bibnamefont
  {Schlosser}}, \bibinfo {author} {\bibfnamefont {G.}~\bibnamefont {Reymond}},
  \bibinfo {author} {\bibfnamefont {I.}~\bibnamefont {Protsenko}}, \ and\
  \bibinfo {author} {\bibfnamefont {P.}~\bibnamefont {Grangier}},\ }\href
  {\doibase 10.1038/35082512} {\bibfield  {journal} {\bibinfo  {journal}
  {Nature}\ }\textbf {\bibinfo {volume} {411}},\ \bibinfo {pages} {1024}
  (\bibinfo {year} {2001})}\BibitemShut {NoStop}%
\bibitem [{\citenamefont {Kaufman}\ and\ \citenamefont
  {Ni}(2021)}]{kaufman2021quantum}%
  \BibitemOpen
  \bibfield  {author} {\bibinfo {author} {\bibfnamefont {A.~M.}\ \bibnamefont
  {Kaufman}}\ and\ \bibinfo {author} {\bibfnamefont {K.-K.}\ \bibnamefont
  {Ni}},\ }\href {\doibase 10.1038/s41567-021-01357-2} {\bibfield  {journal}
  {\bibinfo  {journal} {Nature Physics}\ }\textbf {\bibinfo {volume} {17}},\
  \bibinfo {pages} {1324} (\bibinfo {year} {2021})}\BibitemShut {NoStop}%
\bibitem [{\citenamefont {Kim}\ \emph {et~al.}(2019)\citenamefont {Kim},
  \citenamefont {Chang}, \citenamefont {Fields}, \citenamefont {Chen},\ and\
  \citenamefont {Hung}}]{Kim2019NC}%
  \BibitemOpen
  \bibfield  {author} {\bibinfo {author} {\bibfnamefont {M.~E.}\ \bibnamefont
  {Kim}}, \bibinfo {author} {\bibfnamefont {T.-H.}\ \bibnamefont {Chang}},
  \bibinfo {author} {\bibfnamefont {B.~M.}\ \bibnamefont {Fields}}, \bibinfo
  {author} {\bibfnamefont {C.-A.}\ \bibnamefont {Chen}}, \ and\ \bibinfo
  {author} {\bibfnamefont {C.-L.}\ \bibnamefont {Hung}},\ }\href {\doibase
  10.1038/s41467-019-09635-7} {\bibfield  {journal} {\bibinfo  {journal}
  {Nature Communications}\ }\textbf {\bibinfo {volume} {10}},\ \bibinfo {pages}
  {1647} (\bibinfo {year} {2019})}\BibitemShut {NoStop}%
\bibitem [{\citenamefont {Nayak}\ \emph {et~al.}(2019)\citenamefont {Nayak},
  \citenamefont {Wang},\ and\ \citenamefont {Keloth}}]{Nayak19}%
  \BibitemOpen
  \bibfield  {author} {\bibinfo {author} {\bibfnamefont {K.~P.}\ \bibnamefont
  {Nayak}}, \bibinfo {author} {\bibfnamefont {J.}~\bibnamefont {Wang}}, \ and\
  \bibinfo {author} {\bibfnamefont {J.}~\bibnamefont {Keloth}},\ }\href
  {\doibase 10.1103/PhysRevLett.123.213602} {\bibfield  {journal} {\bibinfo
  {journal} {Phys. Rev. Lett.}\ }\textbf {\bibinfo {volume} {123}},\ \bibinfo
  {pages} {213602} (\bibinfo {year} {2019})}\BibitemShut {NoStop}%
\bibitem [{\citenamefont {Liu}\ \emph {et~al.}(2022{\natexlab{a}})\citenamefont
  {Liu}, \citenamefont {Liu}, \citenamefont {Peng}, \citenamefont {Xu},
  \citenamefont {Chen}, \citenamefont {Ren}, \citenamefont {Wang},\ and\
  \citenamefont {Zou}}]{liu2022multigrating}%
  \BibitemOpen
  \bibfield  {author} {\bibinfo {author} {\bibfnamefont {A.}~\bibnamefont
  {Liu}}, \bibinfo {author} {\bibfnamefont {J.}~\bibnamefont {Liu}}, \bibinfo
  {author} {\bibfnamefont {W.}~\bibnamefont {Peng}}, \bibinfo {author}
  {\bibfnamefont {X.-B.}\ \bibnamefont {Xu}}, \bibinfo {author} {\bibfnamefont
  {G.-J.}\ \bibnamefont {Chen}}, \bibinfo {author} {\bibfnamefont
  {X.}~\bibnamefont {Ren}}, \bibinfo {author} {\bibfnamefont {Q.}~\bibnamefont
  {Wang}}, \ and\ \bibinfo {author} {\bibfnamefont {C.-L.}\ \bibnamefont
  {Zou}},\ }\href {\doibase 10.1103/PhysRevA.105.053520} {\bibfield  {journal}
  {\bibinfo  {journal} {Physical Review A}\ }\textbf {\bibinfo {volume}
  {105}},\ \bibinfo {pages} {053520} (\bibinfo {year}
  {2022}{\natexlab{a}})}\BibitemShut {NoStop}%
\bibitem [{\citenamefont {Frese}\ \emph {et~al.}(2000)\citenamefont {Frese},
  \citenamefont {Ueberholz}, \citenamefont {Kuhr}, \citenamefont {Alt},
  \citenamefont {Schrader}, \citenamefont {Gomer},\ and\ \citenamefont
  {Meschede}}]{Frese00}%
  \BibitemOpen
  \bibfield  {author} {\bibinfo {author} {\bibfnamefont {D.}~\bibnamefont
  {Frese}}, \bibinfo {author} {\bibfnamefont {B.}~\bibnamefont {Ueberholz}},
  \bibinfo {author} {\bibfnamefont {S.}~\bibnamefont {Kuhr}}, \bibinfo {author}
  {\bibfnamefont {W.}~\bibnamefont {Alt}}, \bibinfo {author} {\bibfnamefont
  {D.}~\bibnamefont {Schrader}}, \bibinfo {author} {\bibfnamefont
  {V.}~\bibnamefont {Gomer}}, \ and\ \bibinfo {author} {\bibfnamefont
  {D.}~\bibnamefont {Meschede}},\ }\href {\doibase 10.1103/PhysRevLett.85.3777}
  {\bibfield  {journal} {\bibinfo  {journal} {Phys. Rev. Lett.}\ }\textbf
  {\bibinfo {volume} {85}},\ \bibinfo {pages} {3777} (\bibinfo {year}
  {2000})}\BibitemShut {NoStop}%
\bibitem [{\citenamefont {Ton}\ \emph {et~al.}(2022)\citenamefont {Ton},
  \citenamefont {Kestler}, \citenamefont {Filin}, \citenamefont {Cheung},
  \citenamefont {Schneeweiss}, \citenamefont {Hoinkes}, \citenamefont {Volz},
  \citenamefont {Safronova}, \citenamefont {Rauschenbeutel},\ and\
  \citenamefont {Barreiro}}]{ton2022state}%
  \BibitemOpen
  \bibfield  {author} {\bibinfo {author} {\bibfnamefont {K.}~\bibnamefont
  {Ton}}, \bibinfo {author} {\bibfnamefont {G.}~\bibnamefont {Kestler}},
  \bibinfo {author} {\bibfnamefont {D.}~\bibnamefont {Filin}}, \bibinfo
  {author} {\bibfnamefont {C.}~\bibnamefont {Cheung}}, \bibinfo {author}
  {\bibfnamefont {P.}~\bibnamefont {Schneeweiss}}, \bibinfo {author}
  {\bibfnamefont {T.}~\bibnamefont {Hoinkes}}, \bibinfo {author} {\bibfnamefont
  {J.}~\bibnamefont {Volz}}, \bibinfo {author} {\bibfnamefont {M.}~\bibnamefont
  {Safronova}}, \bibinfo {author} {\bibfnamefont {A.}~\bibnamefont
  {Rauschenbeutel}}, \ and\ \bibinfo {author} {\bibfnamefont {J.}~\bibnamefont
  {Barreiro}},\ }\href {http://arxiv.org/abs/2211.04004} {\bibfield  {journal}
  {\bibinfo  {journal} {arXiv:~2211.04004}\ } (\bibinfo {year}
  {2022})}\BibitemShut {NoStop}%
\bibitem [{\citenamefont {Raimond}\ \emph {et~al.}(2001)\citenamefont
  {Raimond}, \citenamefont {Brune},\ and\ \citenamefont
  {Haroche}}]{Raimond2001}%
  \BibitemOpen
  \bibfield  {author} {\bibinfo {author} {\bibfnamefont {J.~M.}\ \bibnamefont
  {Raimond}}, \bibinfo {author} {\bibfnamefont {M.}~\bibnamefont {Brune}}, \
  and\ \bibinfo {author} {\bibfnamefont {S.}~\bibnamefont {Haroche}},\ }\href
  {\doibase 10.1103/RevModPhys.73.565} {\bibfield  {journal} {\bibinfo
  {journal} {Reviews of Modern Physics}\ }\textbf {\bibinfo {volume} {73}},\
  \bibinfo {pages} {565} (\bibinfo {year} {2001})}\BibitemShut {NoStop}%
\bibitem [{\citenamefont {Reiserer}\ and\ \citenamefont
  {Rempe}(2015)}]{Reiserer2015}%
  \BibitemOpen
  \bibfield  {author} {\bibinfo {author} {\bibfnamefont {A.}~\bibnamefont
  {Reiserer}}\ and\ \bibinfo {author} {\bibfnamefont {G.}~\bibnamefont
  {Rempe}},\ }\href {\doibase 10.1103/RevModPhys.87.1379} {\bibfield  {journal}
  {\bibinfo  {journal} {Reviews of Modern Physics}\ }\textbf {\bibinfo {volume}
  {87}},\ \bibinfo {pages} {1379} (\bibinfo {year} {2015})}\BibitemShut
  {NoStop}%
\bibitem [{\citenamefont {Liu}\ \emph {et~al.}(2022{\natexlab{b}})\citenamefont
  {Liu}, \citenamefont {Wang}, \citenamefont {Yang}, \citenamefont {Wang},
  \citenamefont {Fan}, \citenamefont {Li}, \citenamefont {Zhang},\ and\
  \citenamefont {Zhang}}]{Liu2022SXU}%
  \BibitemOpen
  \bibfield  {author} {\bibinfo {author} {\bibfnamefont {Y.}~\bibnamefont
  {Liu}}, \bibinfo {author} {\bibfnamefont {Z.}~\bibnamefont {Wang}}, \bibinfo
  {author} {\bibfnamefont {P.}~\bibnamefont {Yang}}, \bibinfo {author}
  {\bibfnamefont {Q.}~\bibnamefont {Wang}}, \bibinfo {author} {\bibfnamefont
  {Q.}~\bibnamefont {Fan}}, \bibinfo {author} {\bibfnamefont {G.}~\bibnamefont
  {Li}}, \bibinfo {author} {\bibfnamefont {P.}~\bibnamefont {Zhang}}, \ and\
  \bibinfo {author} {\bibfnamefont {T.}~\bibnamefont {Zhang}},\ }\href
  {http://arxiv.org/abs/2207.04371} {\bibfield  {journal} {\bibinfo  {journal}
  {arXiv: 2207.04371}\ } (\bibinfo {year} {2022}{\natexlab{b}})}\BibitemShut
  {NoStop}%
\bibitem [{\citenamefont {Chen}\ \emph
  {et~al.}(2022{\natexlab{a}})\citenamefont {Chen}, \citenamefont {Szurek},
  \citenamefont {Hu}, \citenamefont {de~Hond}, \citenamefont {Braverman},\ and\
  \citenamefont {Vuletic}}]{Chen2022}%
  \BibitemOpen
  \bibfield  {author} {\bibinfo {author} {\bibfnamefont {Y.-T.}\ \bibnamefont
  {Chen}}, \bibinfo {author} {\bibfnamefont {M.}~\bibnamefont {Szurek}},
  \bibinfo {author} {\bibfnamefont {B.}~\bibnamefont {Hu}}, \bibinfo {author}
  {\bibfnamefont {J.}~\bibnamefont {de~Hond}}, \bibinfo {author} {\bibfnamefont
  {B.}~\bibnamefont {Braverman}}, \ and\ \bibinfo {author} {\bibfnamefont
  {V.}~\bibnamefont {Vuletic}},\ }\href {\doibase 10.1364/OE.469644} {\bibfield
   {journal} {\bibinfo  {journal} {Optics Express}\ }\textbf {\bibinfo {volume}
  {30}},\ \bibinfo {pages} {37426} (\bibinfo {year}
  {2022}{\natexlab{a}})}\BibitemShut {NoStop}%
\bibitem [{\citenamefont {Deist}\ \emph {et~al.}(2022)\citenamefont {Deist},
  \citenamefont {Lu}, \citenamefont {Ho}, \citenamefont {Pasha}, \citenamefont
  {Zeiher}, \citenamefont {Yan},\ and\ \citenamefont
  {Stamper-Kurn}}]{Deist2022}%
  \BibitemOpen
  \bibfield  {author} {\bibinfo {author} {\bibfnamefont {E.}~\bibnamefont
  {Deist}}, \bibinfo {author} {\bibfnamefont {Y.-H.}\ \bibnamefont {Lu}},
  \bibinfo {author} {\bibfnamefont {J.}~\bibnamefont {Ho}}, \bibinfo {author}
  {\bibfnamefont {M.~K.}\ \bibnamefont {Pasha}}, \bibinfo {author}
  {\bibfnamefont {J.}~\bibnamefont {Zeiher}}, \bibinfo {author} {\bibfnamefont
  {Z.}~\bibnamefont {Yan}}, \ and\ \bibinfo {author} {\bibfnamefont {D.~M.}\
  \bibnamefont {Stamper-Kurn}},\ }\href {\doibase
  10.1103/PhysRevLett.129.203602} {\bibfield  {journal} {\bibinfo  {journal}
  {Physical Review Letters}\ }\textbf {\bibinfo {volume} {129}},\ \bibinfo
  {pages} {203602} (\bibinfo {year} {2022})}\BibitemShut {NoStop}%
\bibitem [{\citenamefont {Daiss}\ \emph {et~al.}(2021)\citenamefont {Daiss},
  \citenamefont {Langenfeld}, \citenamefont {Welte}, \citenamefont {Distante},
  \citenamefont {Thomas}, \citenamefont {Hartung}, \citenamefont {Morin},\ and\
  \citenamefont {Rempe}}]{Daiss2021}%
  \BibitemOpen
  \bibfield  {author} {\bibinfo {author} {\bibfnamefont {S.}~\bibnamefont
  {Daiss}}, \bibinfo {author} {\bibfnamefont {S.}~\bibnamefont {Langenfeld}},
  \bibinfo {author} {\bibfnamefont {S.}~\bibnamefont {Welte}}, \bibinfo
  {author} {\bibfnamefont {E.}~\bibnamefont {Distante}}, \bibinfo {author}
  {\bibfnamefont {P.}~\bibnamefont {Thomas}}, \bibinfo {author} {\bibfnamefont
  {L.}~\bibnamefont {Hartung}}, \bibinfo {author} {\bibfnamefont
  {O.}~\bibnamefont {Morin}}, \ and\ \bibinfo {author} {\bibfnamefont
  {G.}~\bibnamefont {Rempe}},\ }\href {\doibase 10.1126/science.abe3150}
  {\bibfield  {journal} {\bibinfo  {journal} {Science}\ }\textbf {\bibinfo
  {volume} {371}},\ \bibinfo {pages} {614} (\bibinfo {year} {2021})},\ \Eprint
  {http://arxiv.org/abs/2103.13095} {arXiv:2103.13095} \BibitemShut {NoStop}%
\bibitem [{\citenamefont {Reiserer}(2022)}]{Reiserer2022}%
  \BibitemOpen
  \bibfield  {author} {\bibinfo {author} {\bibfnamefont {A.}~\bibnamefont
  {Reiserer}},\ }\href {\doibase 10.1103/RevModPhys.94.041003} {\bibfield
  {journal} {\bibinfo  {journal} {Reviews of Modern Physics}\ }\textbf
  {\bibinfo {volume} {94}},\ \bibinfo {pages} {041003} (\bibinfo {year}
  {2022})},\ \Eprint {http://arxiv.org/abs/2205.15380} {arXiv:2205.15380}
  \BibitemShut {NoStop}%
\bibitem [{\citenamefont {Chang}\ \emph {et~al.}(2019)\citenamefont {Chang},
  \citenamefont {Fields}, \citenamefont {Kim},\ and\ \citenamefont
  {Hung}}]{chang2019microring}%
  \BibitemOpen
  \bibfield  {author} {\bibinfo {author} {\bibfnamefont {T.-H.}\ \bibnamefont
  {Chang}}, \bibinfo {author} {\bibfnamefont {B.~M.}\ \bibnamefont {Fields}},
  \bibinfo {author} {\bibfnamefont {M.~E.}\ \bibnamefont {Kim}}, \ and\
  \bibinfo {author} {\bibfnamefont {C.-L.}\ \bibnamefont {Hung}},\ }\href@noop
  {} {\bibfield  {journal} {\bibinfo  {journal} {Optica}\ }\textbf {\bibinfo
  {volume} {6}},\ \bibinfo {pages} {1203} (\bibinfo {year} {2019})}\BibitemShut
  {NoStop}%
\bibitem [{\citenamefont {B{\'e}guin}\ \emph {et~al.}(2020)\citenamefont
  {B{\'e}guin}, \citenamefont {Burgers}, \citenamefont {Luan}, \citenamefont
  {Qin}, \citenamefont {Yu},\ and\ \citenamefont
  {Kimble}}]{beguin2020advanced}%
  \BibitemOpen
  \bibfield  {author} {\bibinfo {author} {\bibfnamefont {J.-B.}\ \bibnamefont
  {B{\'e}guin}}, \bibinfo {author} {\bibfnamefont {A.}~\bibnamefont {Burgers}},
  \bibinfo {author} {\bibfnamefont {X.}~\bibnamefont {Luan}}, \bibinfo {author}
  {\bibfnamefont {Z.}~\bibnamefont {Qin}}, \bibinfo {author} {\bibfnamefont
  {S.-P.}\ \bibnamefont {Yu}}, \ and\ \bibinfo {author} {\bibfnamefont {H.~J.}\
  \bibnamefont {Kimble}},\ }\href {\doibase 10.1364/OPTICA.384408} {\bibfield
  {journal} {\bibinfo  {journal} {Optica}\ }\textbf {\bibinfo {volume} {7}},\
  \bibinfo {pages} {1} (\bibinfo {year} {2020})}\BibitemShut {NoStop}%
\bibitem [{\citenamefont {Luan}\ \emph {et~al.}(2020)\citenamefont {Luan},
  \citenamefont {B{\'e}guin}, \citenamefont {Burgers}, \citenamefont {Qin},
  \citenamefont {Yu},\ and\ \citenamefont {Kimble}}]{Luan2020tweezers}%
  \BibitemOpen
  \bibfield  {author} {\bibinfo {author} {\bibfnamefont {X.}~\bibnamefont
  {Luan}}, \bibinfo {author} {\bibfnamefont {J.-B.}\ \bibnamefont
  {B{\'e}guin}}, \bibinfo {author} {\bibfnamefont {A.~P.}\ \bibnamefont
  {Burgers}}, \bibinfo {author} {\bibfnamefont {Z.}~\bibnamefont {Qin}},
  \bibinfo {author} {\bibfnamefont {S.-P.}\ \bibnamefont {Yu}}, \ and\ \bibinfo
  {author} {\bibfnamefont {H.~J.}\ \bibnamefont {Kimble}},\ }\href {\doibase
  10.1002/qute.202000008} {\bibfield  {journal} {\bibinfo  {journal} {Advanced
  Quantum Technologies}\ }\textbf {\bibinfo {volume} {3}},\ \bibinfo {pages}
  {2000008} (\bibinfo {year} {2020})}\BibitemShut {NoStop}%
\bibitem [{\citenamefont {Liu}\ \emph {et~al.}(2022{\natexlab{c}})\citenamefont
  {Liu}, \citenamefont {Xu}, \citenamefont {Xu}, \citenamefont {Chen},
  \citenamefont {Zhang}, \citenamefont {Xiang}, \citenamefont {Guo},
  \citenamefont {Wang},\ and\ \citenamefont {Zou}}]{liu2022proposal}%
  \BibitemOpen
  \bibfield  {author} {\bibinfo {author} {\bibfnamefont {A.}~\bibnamefont
  {Liu}}, \bibinfo {author} {\bibfnamefont {L.}~\bibnamefont {Xu}}, \bibinfo
  {author} {\bibfnamefont {X.-B.}\ \bibnamefont {Xu}}, \bibinfo {author}
  {\bibfnamefont {G.-J.}\ \bibnamefont {Chen}}, \bibinfo {author}
  {\bibfnamefont {P.}~\bibnamefont {Zhang}}, \bibinfo {author} {\bibfnamefont
  {G.-Y.}\ \bibnamefont {Xiang}}, \bibinfo {author} {\bibfnamefont {G.-C.}\
  \bibnamefont {Guo}}, \bibinfo {author} {\bibfnamefont {Q.}~\bibnamefont
  {Wang}}, \ and\ \bibinfo {author} {\bibfnamefont {C.-L.}\ \bibnamefont
  {Zou}},\ }\href {\doibase 10.1103/PhysRevA.106.033104} {\bibfield  {journal}
  {\bibinfo  {journal} {Physical Review A}\ }\textbf {\bibinfo {volume}
  {106}},\ \bibinfo {pages} {033104} (\bibinfo {year}
  {2022}{\natexlab{c}})}\BibitemShut {NoStop}%
\bibitem [{\citenamefont {Wang}\ \emph {et~al.}(2022)\citenamefont {Wang},
  \citenamefont {Xu},\ and\ \citenamefont {Chai}}]{Wang2022chip}%
  \BibitemOpen
  \bibfield  {author} {\bibinfo {author} {\bibfnamefont {W.}~\bibnamefont
  {Wang}}, \bibinfo {author} {\bibfnamefont {Y.}~\bibnamefont {Xu}}, \ and\
  \bibinfo {author} {\bibfnamefont {Z.}~\bibnamefont {Chai}},\ }\href {\doibase
  10.1002/adpr.202200153} {\bibfield  {journal} {\bibinfo  {journal} {Advanced
  Photonics Research}\ }\textbf {\bibinfo {volume} {3}},\ \bibinfo {pages}
  {2200153} (\bibinfo {year} {2022})}\BibitemShut {NoStop}%
\bibitem [{\citenamefont {Chen}\ \emph
  {et~al.}(2022{\natexlab{b}})\citenamefont {Chen}, \citenamefont {Huang},
  \citenamefont {Xu}, \citenamefont {Zhang}, \citenamefont {Ma}, \citenamefont
  {Lu}, \citenamefont {Wang}, \citenamefont {Chen}, \citenamefont {Zhang},
  \citenamefont {Tang}, \citenamefont {Dong}, \citenamefont {Liu},
  \citenamefont {Xiang}, \citenamefont {Guo},\ and\ \citenamefont
  {Zou}}]{PrAchenliang}%
  \BibitemOpen
  \bibfield  {author} {\bibinfo {author} {\bibfnamefont {L.}~\bibnamefont
  {Chen}}, \bibinfo {author} {\bibfnamefont {C.-J.}\ \bibnamefont {Huang}},
  \bibinfo {author} {\bibfnamefont {X.-B.}\ \bibnamefont {Xu}}, \bibinfo
  {author} {\bibfnamefont {Y.-C.}\ \bibnamefont {Zhang}}, \bibinfo {author}
  {\bibfnamefont {D.-Q.}\ \bibnamefont {Ma}}, \bibinfo {author} {\bibfnamefont
  {Z.-T.}\ \bibnamefont {Lu}}, \bibinfo {author} {\bibfnamefont {Z.-B.}\
  \bibnamefont {Wang}}, \bibinfo {author} {\bibfnamefont {G.-J.}\ \bibnamefont
  {Chen}}, \bibinfo {author} {\bibfnamefont {J.-Z.}\ \bibnamefont {Zhang}},
  \bibinfo {author} {\bibfnamefont {H.~X.}\ \bibnamefont {Tang}}, \bibinfo
  {author} {\bibfnamefont {C.-H.}\ \bibnamefont {Dong}}, \bibinfo {author}
  {\bibfnamefont {W.}~\bibnamefont {Liu}}, \bibinfo {author} {\bibfnamefont
  {G.-Y.}\ \bibnamefont {Xiang}}, \bibinfo {author} {\bibfnamefont {G.-C.}\
  \bibnamefont {Guo}}, \ and\ \bibinfo {author} {\bibfnamefont {C.-L.}\
  \bibnamefont {Zou}},\ }\href {\doibase 10.1103/PhysRevApplied.17.034031}
  {\bibfield  {journal} {\bibinfo  {journal} {Phys. Rev. Appl.}\ }\textbf
  {\bibinfo {volume} {17}},\ \bibinfo {pages} {034031} (\bibinfo {year}
  {2022}{\natexlab{b}})}\BibitemShut {NoStop}%
\bibitem [{\citenamefont {Dyer}\ \emph {et~al.}(2022)\citenamefont {Dyer},
  \citenamefont {Gallacher}, \citenamefont {Hawley}, \citenamefont {Bregazzi},
  \citenamefont {Griffin}, \citenamefont {Arnold}, \citenamefont {Paul},
  \citenamefont {Riis},\ and\ \citenamefont {McGilligan}}]{Dyer2022}%
  \BibitemOpen
  \bibfield  {author} {\bibinfo {author} {\bibfnamefont {S.}~\bibnamefont
  {Dyer}}, \bibinfo {author} {\bibfnamefont {K.}~\bibnamefont {Gallacher}},
  \bibinfo {author} {\bibfnamefont {U.}~\bibnamefont {Hawley}}, \bibinfo
  {author} {\bibfnamefont {A.}~\bibnamefont {Bregazzi}}, \bibinfo {author}
  {\bibfnamefont {P.~F.}\ \bibnamefont {Griffin}}, \bibinfo {author}
  {\bibfnamefont {A.~S.}\ \bibnamefont {Arnold}}, \bibinfo {author}
  {\bibfnamefont {D.~J.}\ \bibnamefont {Paul}}, \bibinfo {author}
  {\bibfnamefont {E.}~\bibnamefont {Riis}}, \ and\ \bibinfo {author}
  {\bibfnamefont {J.~P.}\ \bibnamefont {McGilligan}},\ }\href
  {http://arxiv.org/abs/2212.02853} {\bibfield  {journal} {\bibinfo  {journal}
  {arXiv:~2212.02853}\ } (\bibinfo {year} {2022})}\BibitemShut {NoStop}%
\bibitem [{\citenamefont {Isichenko}\ \emph {et~al.}(2022)\citenamefont
  {Isichenko}, \citenamefont {Chauhan}, \citenamefont {Bose}, \citenamefont
  {Wang}, \citenamefont {Kunz},\ and\ \citenamefont
  {Blumenthal}}]{Isichenko2022}%
  \BibitemOpen
  \bibfield  {author} {\bibinfo {author} {\bibfnamefont {A.}~\bibnamefont
  {Isichenko}}, \bibinfo {author} {\bibfnamefont {N.}~\bibnamefont {Chauhan}},
  \bibinfo {author} {\bibfnamefont {D.}~\bibnamefont {Bose}}, \bibinfo {author}
  {\bibfnamefont {J.}~\bibnamefont {Wang}}, \bibinfo {author} {\bibfnamefont
  {P.~D.}\ \bibnamefont {Kunz}}, \ and\ \bibinfo {author} {\bibfnamefont
  {D.~J.}\ \bibnamefont {Blumenthal}},\ }\href
  {http://arxiv.org/abs/2212.11417} {\bibfield  {journal} {\bibinfo  {journal}
  {arXiv:~2212.11417}\ } (\bibinfo {year} {2022})}\BibitemShut {NoStop}%
\bibitem [{\citenamefont {Zhou}\ \emph {et~al.}(2023)\citenamefont {Zhou},
  \citenamefont {Tamura}, \citenamefont {Chang},\ and\ \citenamefont
  {Hung}}]{Zhou2023}%
  \BibitemOpen
  \bibfield  {author} {\bibinfo {author} {\bibfnamefont {X.}~\bibnamefont
  {Zhou}}, \bibinfo {author} {\bibfnamefont {H.}~\bibnamefont {Tamura}},
  \bibinfo {author} {\bibfnamefont {T.-H.}\ \bibnamefont {Chang}}, \ and\
  \bibinfo {author} {\bibfnamefont {C.-L.}\ \bibnamefont {Hung}},\ }\href
  {\doibase 10.1103/PhysRevLett.130.103601} {\bibfield  {journal} {\bibinfo
  {journal} {Physical Review Letters}\ }\textbf {\bibinfo {volume} {130}},\
  \bibinfo {pages} {103601} (\bibinfo {year} {2023})}\BibitemShut {NoStop}%
\bibitem [{\citenamefont {Bouscal}\ \emph {et~al.}(2023)\citenamefont
  {Bouscal}, \citenamefont {Kemiche}, \citenamefont {Mahapatra}, \citenamefont
  {Fayard}, \citenamefont {Berroir}, \citenamefont {Ray}, \citenamefont
  {Greffet}, \citenamefont {Raineri}, \citenamefont {Levenson}, \citenamefont
  {Bencheikh}, \citenamefont {Sauvan}, \citenamefont {Urvoy},\ and\
  \citenamefont {Laurat}}]{Bouscal2023}%
  \BibitemOpen
  \bibfield  {author} {\bibinfo {author} {\bibfnamefont {A.}~\bibnamefont
  {Bouscal}}, \bibinfo {author} {\bibfnamefont {M.}~\bibnamefont {Kemiche}},
  \bibinfo {author} {\bibfnamefont {S.}~\bibnamefont {Mahapatra}}, \bibinfo
  {author} {\bibfnamefont {N.}~\bibnamefont {Fayard}}, \bibinfo {author}
  {\bibfnamefont {J.}~\bibnamefont {Berroir}}, \bibinfo {author} {\bibfnamefont
  {T.}~\bibnamefont {Ray}}, \bibinfo {author} {\bibfnamefont {J.-J.}\
  \bibnamefont {Greffet}}, \bibinfo {author} {\bibfnamefont {F.}~\bibnamefont
  {Raineri}}, \bibinfo {author} {\bibfnamefont {A.}~\bibnamefont {Levenson}},
  \bibinfo {author} {\bibfnamefont {K.}~\bibnamefont {Bencheikh}}, \bibinfo
  {author} {\bibfnamefont {C.}~\bibnamefont {Sauvan}}, \bibinfo {author}
  {\bibfnamefont {A.}~\bibnamefont {Urvoy}}, \ and\ \bibinfo {author}
  {\bibfnamefont {J.}~\bibnamefont {Laurat}},\ }\href
  {http://arxiv.org/abs/2301.04675} {\bibfield  {journal} {\bibinfo  {journal}
  {arXiv: 2301.04675}\ } (\bibinfo {year} {2023})}\BibitemShut {NoStop}%
\bibitem [{\citenamefont {Kuhr}\ \emph {et~al.}(2001)\citenamefont {Kuhr},
  \citenamefont {Alt}, \citenamefont {Schrader}, \citenamefont {Muller},
  \citenamefont {Gomer},\ and\ \citenamefont
  {Meschede}}]{kuhr2001deterministic}%
  \BibitemOpen
  \bibfield  {author} {\bibinfo {author} {\bibfnamefont {S.}~\bibnamefont
  {Kuhr}}, \bibinfo {author} {\bibfnamefont {W.}~\bibnamefont {Alt}}, \bibinfo
  {author} {\bibfnamefont {D.}~\bibnamefont {Schrader}}, \bibinfo {author}
  {\bibfnamefont {M.}~\bibnamefont {Muller}}, \bibinfo {author} {\bibfnamefont
  {V.}~\bibnamefont {Gomer}}, \ and\ \bibinfo {author} {\bibfnamefont
  {D.}~\bibnamefont {Meschede}},\ }\href {\doibase 10.1126/science.1062725}
  {\bibfield  {journal} {\bibinfo  {journal} {Science}\ }\textbf {\bibinfo
  {volume} {293}},\ \bibinfo {pages} {278} (\bibinfo {year}
  {2001})}\BibitemShut {NoStop}%
\bibitem [{\citenamefont {Thompson}\ \emph {et~al.}(2013)\citenamefont
  {Thompson}, \citenamefont {Tiecke}, \citenamefont {de~Leon}, \citenamefont
  {Feist}, \citenamefont {Akimov}, \citenamefont {Gullans}, \citenamefont
  {Zibrov}, \citenamefont {Vuleti{\'c}},\ and\ \citenamefont
  {Lukin}}]{thompson2013coupling}%
  \BibitemOpen
  \bibfield  {author} {\bibinfo {author} {\bibfnamefont {J.~D.}\ \bibnamefont
  {Thompson}}, \bibinfo {author} {\bibfnamefont {T.}~\bibnamefont {Tiecke}},
  \bibinfo {author} {\bibfnamefont {N.~P.}\ \bibnamefont {de~Leon}}, \bibinfo
  {author} {\bibfnamefont {J.}~\bibnamefont {Feist}}, \bibinfo {author}
  {\bibfnamefont {A.}~\bibnamefont {Akimov}}, \bibinfo {author} {\bibfnamefont
  {M.}~\bibnamefont {Gullans}}, \bibinfo {author} {\bibfnamefont {A.~S.}\
  \bibnamefont {Zibrov}}, \bibinfo {author} {\bibfnamefont {V.}~\bibnamefont
  {Vuleti{\'c}}}, \ and\ \bibinfo {author} {\bibfnamefont {M.~D.}\ \bibnamefont
  {Lukin}},\ }\href {\doibase 10.1126/science.1237125} {\bibfield  {journal}
  {\bibinfo  {journal} {Science}\ }\textbf {\bibinfo {volume} {340}},\ \bibinfo
  {pages} {1202} (\bibinfo {year} {2013})}\BibitemShut {NoStop}%
\bibitem [{\citenamefont {Burgers}\ \emph {et~al.}(2019)\citenamefont
  {Burgers}, \citenamefont {Peng}, \citenamefont {Muniz}, \citenamefont
  {McClung}, \citenamefont {Martin},\ and\ \citenamefont
  {Kimble}}]{burgers2019clocked}%
  \BibitemOpen
  \bibfield  {author} {\bibinfo {author} {\bibfnamefont {A.~P.}\ \bibnamefont
  {Burgers}}, \bibinfo {author} {\bibfnamefont {L.~S.}\ \bibnamefont {Peng}},
  \bibinfo {author} {\bibfnamefont {J.~A.}\ \bibnamefont {Muniz}}, \bibinfo
  {author} {\bibfnamefont {A.~C.}\ \bibnamefont {McClung}}, \bibinfo {author}
  {\bibfnamefont {M.~J.}\ \bibnamefont {Martin}}, \ and\ \bibinfo {author}
  {\bibfnamefont {H.~J.}\ \bibnamefont {Kimble}},\ }\href {\doibase
  10.1073/pnas.1817249115} {\bibfield  {journal} {\bibinfo  {journal}
  {Proceedings of the National Academy of Sciences}\ }\textbf {\bibinfo
  {volume} {116}},\ \bibinfo {pages} {456} (\bibinfo {year}
  {2019})}\BibitemShut {NoStop}%
\bibitem [{\citenamefont {Will}\ \emph {et~al.}(2021)\citenamefont {Will},
  \citenamefont {Masters}, \citenamefont {Rauschenbeutel}, \citenamefont
  {Scheucher},\ and\ \citenamefont {Volz}}]{Will2021}%
  \BibitemOpen
  \bibfield  {author} {\bibinfo {author} {\bibfnamefont {E.}~\bibnamefont
  {Will}}, \bibinfo {author} {\bibfnamefont {L.}~\bibnamefont {Masters}},
  \bibinfo {author} {\bibfnamefont {A.}~\bibnamefont {Rauschenbeutel}},
  \bibinfo {author} {\bibfnamefont {M.}~\bibnamefont {Scheucher}}, \ and\
  \bibinfo {author} {\bibfnamefont {J.}~\bibnamefont {Volz}},\ }\href {\doibase
  10.1103/PhysRevLett.126.233602} {\bibfield  {journal} {\bibinfo  {journal}
  {Physical Review Letters}\ }\textbf {\bibinfo {volume} {126}},\ \bibinfo
  {pages} {233602} (\bibinfo {year} {2021})},\ \Eprint
  {http://arxiv.org/abs/2010.07267} {arXiv:2010.07267} \BibitemShut {NoStop}%
\bibitem [{\citenamefont {Jannasch}\ \emph {et~al.}(2012)\citenamefont
  {Jannasch}, \citenamefont {Demir{\"o}rs}, \citenamefont {Van~Oostrum},
  \citenamefont {Van~Blaaderen},\ and\ \citenamefont
  {Sch{\"a}ffer}}]{jannasch2012nanonewton}%
  \BibitemOpen
  \bibfield  {author} {\bibinfo {author} {\bibfnamefont {A.}~\bibnamefont
  {Jannasch}}, \bibinfo {author} {\bibfnamefont {A.~F.}\ \bibnamefont
  {Demir{\"o}rs}}, \bibinfo {author} {\bibfnamefont {P.~D.}\ \bibnamefont
  {Van~Oostrum}}, \bibinfo {author} {\bibfnamefont {A.}~\bibnamefont
  {Van~Blaaderen}}, \ and\ \bibinfo {author} {\bibfnamefont {E.}~\bibnamefont
  {Sch{\"a}ffer}},\ }\href {\doibase doi.org/10.1038/nphoton.2012.140}
  {\bibfield  {journal} {\bibinfo  {journal} {Nature Photonics}\ }\textbf
  {\bibinfo {volume} {6}},\ \bibinfo {pages} {469} (\bibinfo {year}
  {2012})}\BibitemShut {NoStop}%
\bibitem [{\citenamefont {Cheng}\ \emph {et~al.}(2016)\citenamefont {Cheng},
  \citenamefont {Song}, \citenamefont {Zhang}, \citenamefont {Jiao},
  \citenamefont {Ma}, \citenamefont {Sui},\ and\ \citenamefont
  {Wang}}]{cheng2016optimal}%
  \BibitemOpen
  \bibfield  {author} {\bibinfo {author} {\bibfnamefont {X.}~\bibnamefont
  {Cheng}}, \bibinfo {author} {\bibfnamefont {Z.}~\bibnamefont {Song}},
  \bibinfo {author} {\bibfnamefont {J.}~\bibnamefont {Zhang}}, \bibinfo
  {author} {\bibfnamefont {H.}~\bibnamefont {Jiao}}, \bibinfo {author}
  {\bibfnamefont {B.}~\bibnamefont {Ma}}, \bibinfo {author} {\bibfnamefont
  {Z.}~\bibnamefont {Sui}}, \ and\ \bibinfo {author} {\bibfnamefont
  {Z.}~\bibnamefont {Wang}},\ }\href {\doibase doi.org/10.1364/OE.24.024313}
  {\bibfield  {journal} {\bibinfo  {journal} {Optics Express}\ }\textbf
  {\bibinfo {volume} {24}},\ \bibinfo {pages} {24313} (\bibinfo {year}
  {2016})}\BibitemShut {NoStop}%
\bibitem [{\citenamefont {Hoenig}\ and\ \citenamefont
  {Moebius}(1991)}]{hoenig1991}%
  \BibitemOpen
  \bibfield  {author} {\bibinfo {author} {\bibfnamefont {D.}~\bibnamefont
  {Hoenig}}\ and\ \bibinfo {author} {\bibfnamefont {D.}~\bibnamefont
  {Moebius}},\ }\href {\doibase doi.org/10.1021/j100165a003} {\bibfield
  {journal} {\bibinfo  {journal} {The Journal of Physical Chemistry}\ }\textbf
  {\bibinfo {volume} {95}},\ \bibinfo {pages} {4590} (\bibinfo {year}
  {1991})}\BibitemShut {NoStop}%
\bibitem [{\citenamefont {Barnett}\ \emph {et~al.}(2000)\citenamefont
  {Barnett}, \citenamefont {Smith}, \citenamefont {Olshanii}, \citenamefont
  {Johnson}, \citenamefont {Adams},\ and\ \citenamefont
  {Prentiss}}]{Barnett2000}%
  \BibitemOpen
  \bibfield  {author} {\bibinfo {author} {\bibfnamefont {A.~H.}\ \bibnamefont
  {Barnett}}, \bibinfo {author} {\bibfnamefont {S.~P.}\ \bibnamefont {Smith}},
  \bibinfo {author} {\bibfnamefont {M.}~\bibnamefont {Olshanii}}, \bibinfo
  {author} {\bibfnamefont {K.~S.}\ \bibnamefont {Johnson}}, \bibinfo {author}
  {\bibfnamefont {A.~W.}\ \bibnamefont {Adams}}, \ and\ \bibinfo {author}
  {\bibfnamefont {M.}~\bibnamefont {Prentiss}},\ }\href {\doibase
  10.1103/PhysRevA.61.023608} {\bibfield  {journal} {\bibinfo  {journal}
  {Physical Review A}\ }\textbf {\bibinfo {volume} {61}},\ \bibinfo {pages}
  {023608} (\bibinfo {year} {2000})}\BibitemShut {NoStop}%
\bibitem [{\citenamefont {Hammes}\ \emph {et~al.}(2003)\citenamefont {Hammes},
  \citenamefont {Rychtarik}, \citenamefont {Engeser}, \citenamefont
  {N{\"{a}}gerl},\ and\ \citenamefont {Grimm}}]{Hammes2003}%
  \BibitemOpen
  \bibfield  {author} {\bibinfo {author} {\bibfnamefont {M.}~\bibnamefont
  {Hammes}}, \bibinfo {author} {\bibfnamefont {D.}~\bibnamefont {Rychtarik}},
  \bibinfo {author} {\bibfnamefont {B.}~\bibnamefont {Engeser}}, \bibinfo
  {author} {\bibfnamefont {H.-C.}\ \bibnamefont {N{\"{a}}gerl}}, \ and\
  \bibinfo {author} {\bibfnamefont {R.}~\bibnamefont {Grimm}},\ }\href
  {\doibase 10.1103/PhysRevLett.90.173001} {\bibfield  {journal} {\bibinfo
  {journal} {Physical Review Letters}\ }\textbf {\bibinfo {volume} {90}},\
  \bibinfo {pages} {173001} (\bibinfo {year} {2003})}\BibitemShut {NoStop}%
\bibitem [{\citenamefont {Le~Kien}\ \emph {et~al.}(2004)\citenamefont
  {Le~Kien}, \citenamefont {Balykin},\ and\ \citenamefont
  {Hakuta}}]{le2004atoms}%
  \BibitemOpen
  \bibfield  {author} {\bibinfo {author} {\bibfnamefont {F.}~\bibnamefont
  {Le~Kien}}, \bibinfo {author} {\bibfnamefont {V.~I.}\ \bibnamefont
  {Balykin}}, \ and\ \bibinfo {author} {\bibfnamefont {K.}~\bibnamefont
  {Hakuta}},\ }\href@noop {} {\bibfield  {journal} {\bibinfo  {journal}
  {Physical Review A}\ }\textbf {\bibinfo {volume} {70}},\ \bibinfo {pages}
  {063403} (\bibinfo {year} {2004})}\BibitemShut {NoStop}%
\bibitem [{\citenamefont {Stievater}\ \emph {et~al.}(2016)\citenamefont
  {Stievater}, \citenamefont {Kozak}, \citenamefont {Pruessner}, \citenamefont
  {Mahon}, \citenamefont {Park}, \citenamefont {Rabinovich},\ and\
  \citenamefont {Fatemi}}]{stievater2016modal}%
  \BibitemOpen
  \bibfield  {author} {\bibinfo {author} {\bibfnamefont {T.~H.}\ \bibnamefont
  {Stievater}}, \bibinfo {author} {\bibfnamefont {D.~A.}\ \bibnamefont
  {Kozak}}, \bibinfo {author} {\bibfnamefont {M.~W.}\ \bibnamefont
  {Pruessner}}, \bibinfo {author} {\bibfnamefont {R.}~\bibnamefont {Mahon}},
  \bibinfo {author} {\bibfnamefont {D.}~\bibnamefont {Park}}, \bibinfo {author}
  {\bibfnamefont {W.~S.}\ \bibnamefont {Rabinovich}}, \ and\ \bibinfo {author}
  {\bibfnamefont {F.~K.}\ \bibnamefont {Fatemi}},\ }\href@noop {} {\bibfield
  {journal} {\bibinfo  {journal} {Optical Materials Express}\ }\textbf
  {\bibinfo {volume} {6}},\ \bibinfo {pages} {3826} (\bibinfo {year}
  {2016})}\BibitemShut {NoStop}%
\bibitem [{\citenamefont {B\'{e}guin}\ \emph {et~al.}(2020)\citenamefont
  {B\'{e}guin}, \citenamefont {Burgers}, \citenamefont {Luan}, \citenamefont
  {Qin}, \citenamefont {Yu},\ and\ \citenamefont {Kimble}}]{Beguin2020}%
  \BibitemOpen
  \bibfield  {author} {\bibinfo {author} {\bibfnamefont {J.-B.}\ \bibnamefont
  {B\'{e}guin}}, \bibinfo {author} {\bibfnamefont {A.~P.}\ \bibnamefont
  {Burgers}}, \bibinfo {author} {\bibfnamefont {X.}~\bibnamefont {Luan}},
  \bibinfo {author} {\bibfnamefont {Z.}~\bibnamefont {Qin}}, \bibinfo {author}
  {\bibfnamefont {S.~P.}\ \bibnamefont {Yu}}, \ and\ \bibinfo {author}
  {\bibfnamefont {H.~J.}\ \bibnamefont {Kimble}},\ }\href {\doibase
  10.1364/OPTICA.384408} {\bibfield  {journal} {\bibinfo  {journal} {Optica}\
  }\textbf {\bibinfo {volume} {7}},\ \bibinfo {pages} {1} (\bibinfo {year}
  {2020})}\BibitemShut {NoStop}%
\bibitem [{\citenamefont {Metcalf}\ and\ \citenamefont {van~der
  Straten}(1999)}]{Metcalf1999}%
  \BibitemOpen
  \bibfield  {author} {\bibinfo {author} {\bibfnamefont {H.~J.}\ \bibnamefont
  {Metcalf}}\ and\ \bibinfo {author} {\bibfnamefont {P.}~\bibnamefont {van~der
  Straten}},\ }\href {\doibase 10.1007/978-1-4612-1470-0} {\emph {\bibinfo
  {title} {Laser Cooling and Trapping}}},\ Graduate Texts in Contemporary
  Physics\ (\bibinfo  {publisher} {Springer New York},\ \bibinfo {address} {New
  York, NY},\ \bibinfo {year} {1999})\BibitemShut {NoStop}%
\bibitem [{\citenamefont {Xu}\ \emph {et~al.}(2023)\citenamefont {Xu},
  \citenamefont {Wang}, \citenamefont {Chen}, \citenamefont {Chen},
  \citenamefont {Yang}, \citenamefont {Xu}, \citenamefont {Liu}, \citenamefont
  {Li}, \citenamefont {Guo}, \citenamefont {Zou},\ and\ \citenamefont
  {Xiang}}]{Xulei2023}%
  \BibitemOpen
  \bibfield  {author} {\bibinfo {author} {\bibfnamefont {L.}~\bibnamefont
  {Xu}}, \bibinfo {author} {\bibfnamefont {L.-X.}\ \bibnamefont {Wang}},
  \bibinfo {author} {\bibfnamefont {G.-J.}\ \bibnamefont {Chen}}, \bibinfo
  {author} {\bibfnamefont {L.}~\bibnamefont {Chen}}, \bibinfo {author}
  {\bibfnamefont {Y.-H.}\ \bibnamefont {Yang}}, \bibinfo {author}
  {\bibfnamefont {X.-B.}\ \bibnamefont {Xu}}, \bibinfo {author} {\bibfnamefont
  {A.}~\bibnamefont {Liu}}, \bibinfo {author} {\bibfnamefont {C.-F.}\
  \bibnamefont {Li}}, \bibinfo {author} {\bibfnamefont {G.-C.}\ \bibnamefont
  {Guo}}, \bibinfo {author} {\bibfnamefont {C.-L.}\ \bibnamefont {Zou}}, \ and\
  \bibinfo {author} {\bibfnamefont {G.-Y.}\ \bibnamefont {Xiang}},\ }\href@noop
  {} {\bibfield  {journal} {\bibinfo  {journal} {Submitted}\ } (\bibinfo {year}
  {2023})}\BibitemShut {NoStop}%
\bibitem [{\citenamefont {Kuhr}\ \emph {et~al.}(2003)\citenamefont {Kuhr},
  \citenamefont {Alt}, \citenamefont {Schrader}, \citenamefont {Dotsenko},
  \citenamefont {Miroshnychenko}, \citenamefont {Rosenfeld}, \citenamefont
  {Khudaverdyan}, \citenamefont {Gomer}, \citenamefont {Rauschenbeutel},\ and\
  \citenamefont {Meschede}}]{Kuhr2003}%
  \BibitemOpen
  \bibfield  {author} {\bibinfo {author} {\bibfnamefont {S.}~\bibnamefont
  {Kuhr}}, \bibinfo {author} {\bibfnamefont {W.}~\bibnamefont {Alt}}, \bibinfo
  {author} {\bibfnamefont {D.}~\bibnamefont {Schrader}}, \bibinfo {author}
  {\bibfnamefont {I.}~\bibnamefont {Dotsenko}}, \bibinfo {author}
  {\bibfnamefont {Y.}~\bibnamefont {Miroshnychenko}}, \bibinfo {author}
  {\bibfnamefont {W.}~\bibnamefont {Rosenfeld}}, \bibinfo {author}
  {\bibfnamefont {M.}~\bibnamefont {Khudaverdyan}}, \bibinfo {author}
  {\bibfnamefont {V.}~\bibnamefont {Gomer}}, \bibinfo {author} {\bibfnamefont
  {A.}~\bibnamefont {Rauschenbeutel}}, \ and\ \bibinfo {author} {\bibfnamefont
  {D.}~\bibnamefont {Meschede}},\ }\href {\doibase
  10.1103/PhysRevLett.91.213002} {\bibfield  {journal} {\bibinfo  {journal}
  {Physical Review Letters}\ }\textbf {\bibinfo {volume} {91}},\ \bibinfo
  {pages} {213002} (\bibinfo {year} {2003})},\ \Eprint
  {http://arxiv.org/abs/0304081} {arXiv:0304081 [quant-ph]} \BibitemShut
  {NoStop}%
\bibitem [{\citenamefont {Schneeweiss}\ \emph {et~al.}(2013)\citenamefont
  {Schneeweiss}, \citenamefont {Dawkins}, \citenamefont {Mitsch}, \citenamefont
  {Reitz}, \citenamefont {Vetsch},\ and\ \citenamefont
  {Rauschenbeutel}}]{Schneeweiss2013}%
  \BibitemOpen
  \bibfield  {author} {\bibinfo {author} {\bibfnamefont {P.}~\bibnamefont
  {Schneeweiss}}, \bibinfo {author} {\bibfnamefont {S.~T.}\ \bibnamefont
  {Dawkins}}, \bibinfo {author} {\bibfnamefont {R.}~\bibnamefont {Mitsch}},
  \bibinfo {author} {\bibfnamefont {D.}~\bibnamefont {Reitz}}, \bibinfo
  {author} {\bibfnamefont {E.}~\bibnamefont {Vetsch}}, \ and\ \bibinfo {author}
  {\bibfnamefont {A.}~\bibnamefont {Rauschenbeutel}},\ }\href {\doibase
  10.1007/s00340-012-5268-2} {\bibfield  {journal} {\bibinfo  {journal}
  {Applied Physics B}\ }\textbf {\bibinfo {volume} {110}},\ \bibinfo {pages}
  {279} (\bibinfo {year} {2013})},\ \Eprint {http://arxiv.org/abs/1207.3021}
  {arXiv:1207.3021} \BibitemShut {NoStop}%
\bibitem [{\citenamefont {Miroshnychenko}\ \emph {et~al.}(2006)\citenamefont
  {Miroshnychenko}, \citenamefont {Alt}, \citenamefont {Dotsenko},
  \citenamefont {F{\"{o}}rster}, \citenamefont {Khudaverdyan}, \citenamefont
  {Meschede}, \citenamefont {Schrader},\ and\ \citenamefont
  {Rauschenbeutel}}]{Miroshnychenko2006}%
  \BibitemOpen
  \bibfield  {author} {\bibinfo {author} {\bibfnamefont {Y.}~\bibnamefont
  {Miroshnychenko}}, \bibinfo {author} {\bibfnamefont {W.}~\bibnamefont {Alt}},
  \bibinfo {author} {\bibfnamefont {I.}~\bibnamefont {Dotsenko}}, \bibinfo
  {author} {\bibfnamefont {L.}~\bibnamefont {F{\"{o}}rster}}, \bibinfo {author}
  {\bibfnamefont {M.}~\bibnamefont {Khudaverdyan}}, \bibinfo {author}
  {\bibfnamefont {D.}~\bibnamefont {Meschede}}, \bibinfo {author}
  {\bibfnamefont {D.}~\bibnamefont {Schrader}}, \ and\ \bibinfo {author}
  {\bibfnamefont {A.}~\bibnamefont {Rauschenbeutel}},\ }\href {\doibase
  10.1038/442151a} {\bibfield  {journal} {\bibinfo  {journal} {Nature}\
  }\textbf {\bibinfo {volume} {442}},\ \bibinfo {pages} {151} (\bibinfo {year}
  {2006})}\BibitemShut {NoStop}%
\bibitem [{\citenamefont {McLachlan}(1964)}]{mclachlan1964van}%
  \BibitemOpen
  \bibfield  {author} {\bibinfo {author} {\bibfnamefont {A.}~\bibnamefont
  {McLachlan}},\ }\href {\doibase 10.1080/00268976300101141} {\bibfield
  {journal} {\bibinfo  {journal} {Molecular Physics}\ }\textbf {\bibinfo
  {volume} {7}},\ \bibinfo {pages} {381} (\bibinfo {year} {1964})}\BibitemShut
  {NoStop}%
\bibitem [{\citenamefont {Ra}\ \emph {et~al.}(1973)\citenamefont {Ra},
  \citenamefont {Bertoni},\ and\ \citenamefont {Felsen}}]{ra1973reflection}%
  \BibitemOpen
  \bibfield  {author} {\bibinfo {author} {\bibfnamefont {J.~W.}\ \bibnamefont
  {Ra}}, \bibinfo {author} {\bibfnamefont {H.}~\bibnamefont {Bertoni}}, \ and\
  \bibinfo {author} {\bibfnamefont {L.}~\bibnamefont {Felsen}},\ }\href@noop {}
  {\bibfield  {journal} {\bibinfo  {journal} {SIAM Journal on Applied
  Mathematics}\ }\textbf {\bibinfo {volume} {24}},\ \bibinfo {pages} {396}
  (\bibinfo {year} {1973})}\BibitemShut {NoStop}%
\bibitem [{\citenamefont {Zhang}\ \emph {et~al.}(2023)\citenamefont {Zhang},
  \citenamefont {Kang}, \citenamefont {Guo}, \citenamefont {Li}, \citenamefont
  {Liu}, \citenamefont {Xie}, \citenamefont {Wu}, \citenamefont {Cai},
  \citenamefont {Gong}, \citenamefont {Shi} \emph {et~al.}}]{zhang2023high}%
  \BibitemOpen
  \bibfield  {author} {\bibinfo {author} {\bibfnamefont {J.}~\bibnamefont
  {Zhang}}, \bibinfo {author} {\bibfnamefont {Y.}~\bibnamefont {Kang}},
  \bibinfo {author} {\bibfnamefont {X.}~\bibnamefont {Guo}}, \bibinfo {author}
  {\bibfnamefont {Y.}~\bibnamefont {Li}}, \bibinfo {author} {\bibfnamefont
  {K.}~\bibnamefont {Liu}}, \bibinfo {author} {\bibfnamefont {Y.}~\bibnamefont
  {Xie}}, \bibinfo {author} {\bibfnamefont {H.}~\bibnamefont {Wu}}, \bibinfo
  {author} {\bibfnamefont {D.}~\bibnamefont {Cai}}, \bibinfo {author}
  {\bibfnamefont {J.}~\bibnamefont {Gong}}, \bibinfo {author} {\bibfnamefont
  {Z.}~\bibnamefont {Shi}},  \emph {et~al.},\ }\href {\doibase
  doi.org/10.1038/s41377-023-01109-2} {\bibfield  {journal} {\bibinfo
  {journal} {Light: Science \& Applications}\ }\textbf {\bibinfo {volume}
  {12}},\ \bibinfo {pages} {89} (\bibinfo {year} {2023})}\BibitemShut {NoStop}%
\bibitem [{\citenamefont {Lodahl}\ \emph {et~al.}(2017)\citenamefont {Lodahl},
  \citenamefont {Mahmoodian}, \citenamefont {Stobbe}, \citenamefont
  {Rauschenbeutel}, \citenamefont {Schneeweiss}, \citenamefont {Volz},
  \citenamefont {Pichler},\ and\ \citenamefont {Zoller}}]{lodahl2017chiral}%
  \BibitemOpen
  \bibfield  {author} {\bibinfo {author} {\bibfnamefont {P.}~\bibnamefont
  {Lodahl}}, \bibinfo {author} {\bibfnamefont {S.}~\bibnamefont {Mahmoodian}},
  \bibinfo {author} {\bibfnamefont {S.}~\bibnamefont {Stobbe}}, \bibinfo
  {author} {\bibfnamefont {A.}~\bibnamefont {Rauschenbeutel}}, \bibinfo
  {author} {\bibfnamefont {P.}~\bibnamefont {Schneeweiss}}, \bibinfo {author}
  {\bibfnamefont {J.}~\bibnamefont {Volz}}, \bibinfo {author} {\bibfnamefont
  {H.}~\bibnamefont {Pichler}}, \ and\ \bibinfo {author} {\bibfnamefont
  {P.}~\bibnamefont {Zoller}},\ }\href {\doibase doi.org/10.1038/nature21037}
  {\bibfield  {journal} {\bibinfo  {journal} {Nature}\ }\textbf {\bibinfo
  {volume} {541}},\ \bibinfo {pages} {473} (\bibinfo {year}
  {2017})}\BibitemShut {NoStop}%
\bibitem [{\citenamefont {Junge}\ \emph {et~al.}(2013)\citenamefont {Junge},
  \citenamefont {O'Shea}, \citenamefont {Volz},\ and\ \citenamefont
  {Rauschenbeutel}}]{junge2013strong}%
  \BibitemOpen
  \bibfield  {author} {\bibinfo {author} {\bibfnamefont {C.}~\bibnamefont
  {Junge}}, \bibinfo {author} {\bibfnamefont {D.}~\bibnamefont {O'Shea}},
  \bibinfo {author} {\bibfnamefont {J.}~\bibnamefont {Volz}}, \ and\ \bibinfo
  {author} {\bibfnamefont {A.}~\bibnamefont {Rauschenbeutel}},\ }\href
  {\doibase doi.org/10.1103/PhysRevLett.110.213604} {\bibfield  {journal}
  {\bibinfo  {journal} {Physical review letters}\ }\textbf {\bibinfo {volume}
  {110}},\ \bibinfo {pages} {213604} (\bibinfo {year} {2013})}\BibitemShut
  {NoStop}%
\bibitem [{\citenamefont {Amico}\ \emph {et~al.}(2022)\citenamefont {Amico},
  \citenamefont {Anderson}, \citenamefont {Boshier}, \citenamefont {Brantut},
  \citenamefont {Kwek}, \citenamefont {Minguzzi},\ and\ \citenamefont {von
  Klitzing}}]{Amico2022}%
  \BibitemOpen
  \bibfield  {author} {\bibinfo {author} {\bibfnamefont {L.}~\bibnamefont
  {Amico}}, \bibinfo {author} {\bibfnamefont {D.}~\bibnamefont {Anderson}},
  \bibinfo {author} {\bibfnamefont {M.}~\bibnamefont {Boshier}}, \bibinfo
  {author} {\bibfnamefont {J.-P.}\ \bibnamefont {Brantut}}, \bibinfo {author}
  {\bibfnamefont {L.-C.}\ \bibnamefont {Kwek}}, \bibinfo {author}
  {\bibfnamefont {A.}~\bibnamefont {Minguzzi}}, \ and\ \bibinfo {author}
  {\bibfnamefont {W.}~\bibnamefont {von Klitzing}},\ }\href {\doibase
  10.1103/RevModPhys.94.041001} {\bibfield  {journal} {\bibinfo  {journal}
  {Reviews of Modern Physics}\ }\textbf {\bibinfo {volume} {94}},\ \bibinfo
  {pages} {041001} (\bibinfo {year} {2022})},\ \Eprint
  {http://arxiv.org/abs/2107.08561} {arXiv:2107.08561} \BibitemShut {NoStop}%
\end{thebibliography}%
\end{document}